\documentclass[pre,twocolumn,floatfix,showpacs,amsmath,amssymb]{revtex4}
\usepackage{graphicx}
\begin{document}
\title{Structural Studies of Decaying Fluid Turbulence: Effect of Initial 
Conditions}
\author{Chirag Kalelkar}
\email{kalelkar@physics.iisc.ernet.in}
\affiliation{Centre for Condensed Matter Theory,
Department of Physics, Indian Institute of Science, Bangalore 560012, India.}
\begin{abstract}
We present results from a systematic numerical study of 
structural properties of an unforced, incompressible, 
homogeneous, and isotropic three-dimensional turbulent fluid with 
an initial energy spectrum that develops a cascade of kinetic energy to 
large wavenumbers. 
The results are compared with those from a recently studied 
set of power-law initial energy spectra [C. Kalelkar and R. Pandit, 
Phys. Rev. E, {\bf 69}, 046304 (2004)] which do not exhibit such a cascade. 
Differences are exhibited in plots of vorticity isosurfaces, the temporal 
evolution of the kinetic energy-dissipation 
rate, and the rates of production of the mean enstrophy along the 
principal axes of the strain-rate tensor. A crossover between 
non-`cascade-type' and `cascade-type' behaviour is shown numerically for 
a specific set of initial energy spectra. 
\end{abstract}
\maketitle
\section{Introduction}
Studies of structural properties (in space) of a turbulent fluid may lead to an 
understanding of dynamically significant processes in turbulence. Homogeneous 
and isotropic turbulence, within the context of {\it statistically 
steady} flows, has been the subject of extensive laboratory\cite{Douady} and 
numerical\cite{Siggia,Vincent,Jimenez} studies. 
These studies show that the turbulent fluid has a kinetic energy spectrum 
with a wavenumber range that exhibits a power-law with exponent equal 
to $-5/3$, in 
accordance with a phenomenological theory due to Kolmogorov\cite{Kolm,Monin}. 
In the statistical steady state, visualization of the vorticity field reveals 
that regions of intense vorticity are organized in slender 
filaments\cite{Douady,Siggia,Vincent,Jimenez}. The 
vorticity vector is observed to preferentially align\cite{Ashurst} with the 
eigenvector associated with the intermediate eigenvalue of the strain-rate 
tensor of the fluid. By contrast, systematic 
spatial studies of {\it decaying} turbulence are scarce and restricted 
to numerical studies\cite{Yamamoto,Hosokawa} of vorticity isosurfaces with 
initial energy spectra of the type that develop a cascade of kinetic energy 
to large wavenumbers.

A recent study\cite{Kalelkar}, investigated the decay of unforced, 
incompressible, homogeneous, and isotropic three-dimensional 
magnetohydrodynamic turbulence from power-law initial conditions. The 
study was a generalization of results\cite{Ditlevsen} obtained for the 
corresponding fluid case. In particular, it was shown both analytically and 
numerically, that for the power-law initial energy spectrum $E(k,t_0)\sim k$ 
($k=|{\bf k}|$ is the magnitude of the wave vector and $t=t_0$ is the 
choice of virtual origin of 
time), the kinetic energy $E(t)$ was found to 
decay as $t^{-1}$ and the integral length scale $L(t)$ was found to grow 
as $t^{0.5}$. Such power-law initial conditions are of interest in the 
astrophysical context of the decay of `primordial' energy spectra\cite{Olesen}. It was shown numerically\cite{Kalelkar}, that such an initial 
energy spectrum does not develop a cascade to large wavenumbers; 
however, structural properties had not been probed.

In this paper, we present results from a pseudospectral direct-numerical 
simulation (DNS) of the unforced, incompressible, three-dimensional 
Navier-Stokes equations and contrast spatial results obtained from an 
initial energy spectrum 
that develops a cascade (Section II), with results from the power-law initial 
spectrum (Section III) mentioned above. Plots of vorticity isosurfaces are 
found to differ, 
with distinct filaments in regions of intense vorticity, 
lacking in the case with power-law initial spectrum. The 
temporal evolution of the kinetic energy-dissipation rate and the rates of 
production 
of the mean enstrophy along the principal axes of the strain-rate tensor, are 
also found to differ. 
However, preferential alignment of the vorticity vector with the intermediate 
eigenvector of the strain-rate tensor is found to be unchanged.  We also 
compare some classical 
results on mean enstrophy production in decaying turbulence. We show 
numerically that a crossover 
between the above-mentioned cases may be realised with an appropriate choice of 
initial conditions.

The Navier-Stokes equations in vorticity form are
\begin{eqnarray}
&&\frac{D\omega_i}{Dt}=S_{ij}\omega_j+\nu\frac{\partial^2\omega_i}
{\partial x_j\partial x_j},
\label{nse}
\end{eqnarray}
where $D/Dt\equiv\partial_t+v_j\partial_j$ is the 
material derivative, $\nu$ is the kinematic viscosity, 
$S_{ij}\equiv1/2(\partial_jv_i+\partial_iv_j)$ is the strain-rate tensor, and 
$\omega_i\equiv\epsilon_{ijk}\partial_jv_k$ is the vorticity ($\epsilon_{ijk}$ 
is the Levi-Civita tensor),  $i,j,k=1,2,3$, with a summation implicit over 
repeated indices. We enforce the incompressibility condition $\partial_iv_i=0$.
\section{Cascade-Type Spectrum}
\subsection{Numerical Method}
\begin{figure}
\includegraphics[height=2.0in]{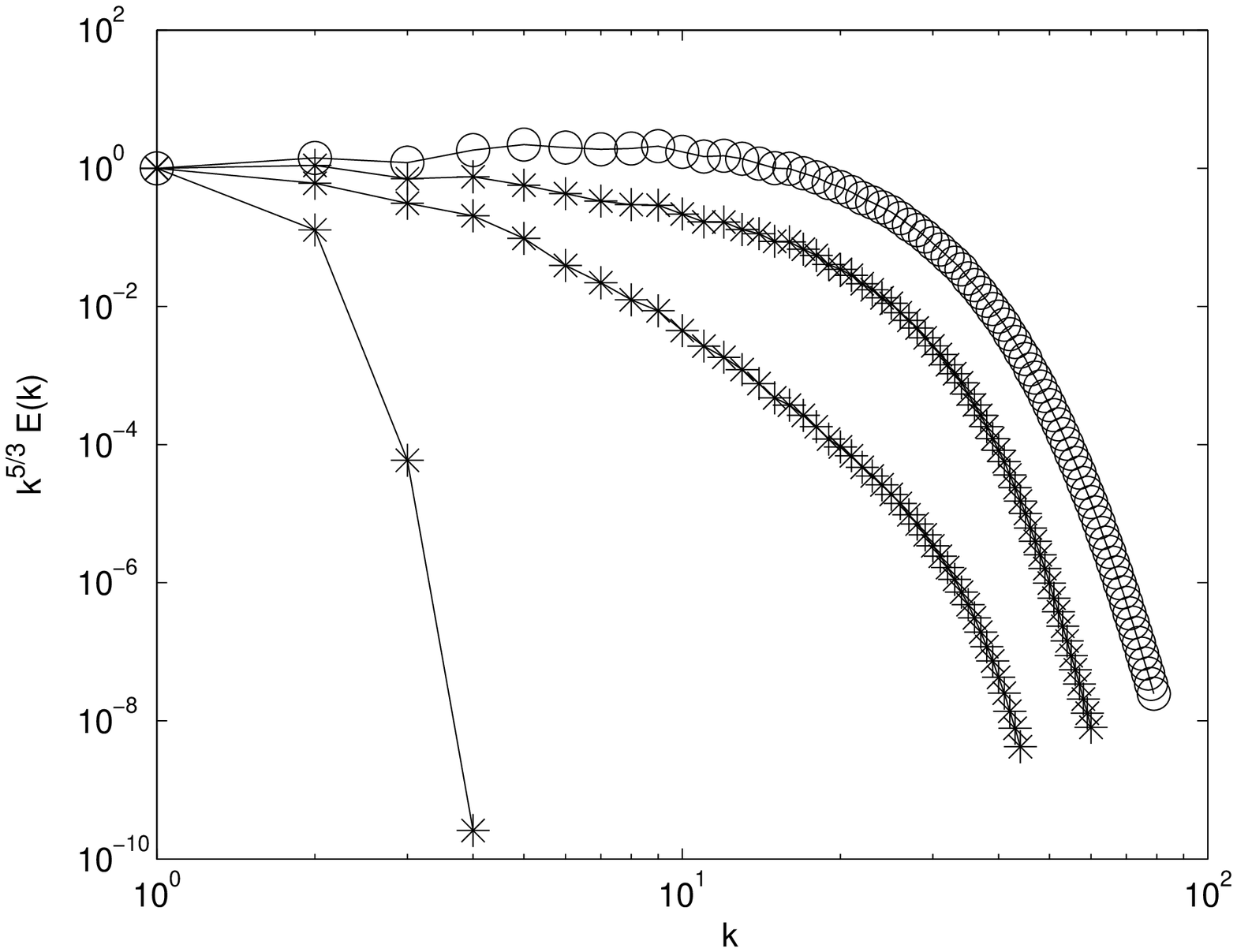}
\includegraphics[height=2.0in]{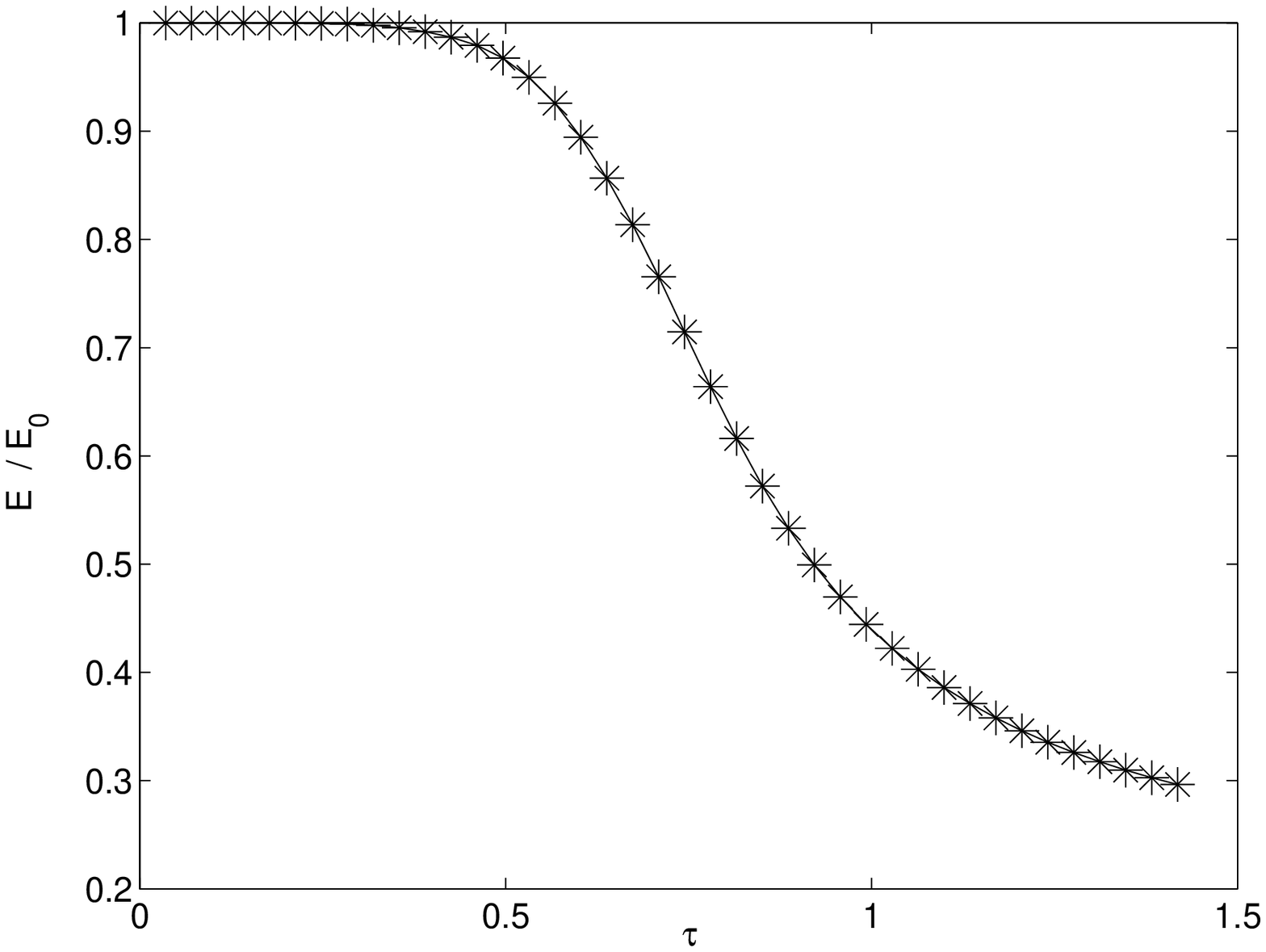}
\includegraphics[height=2.0in]{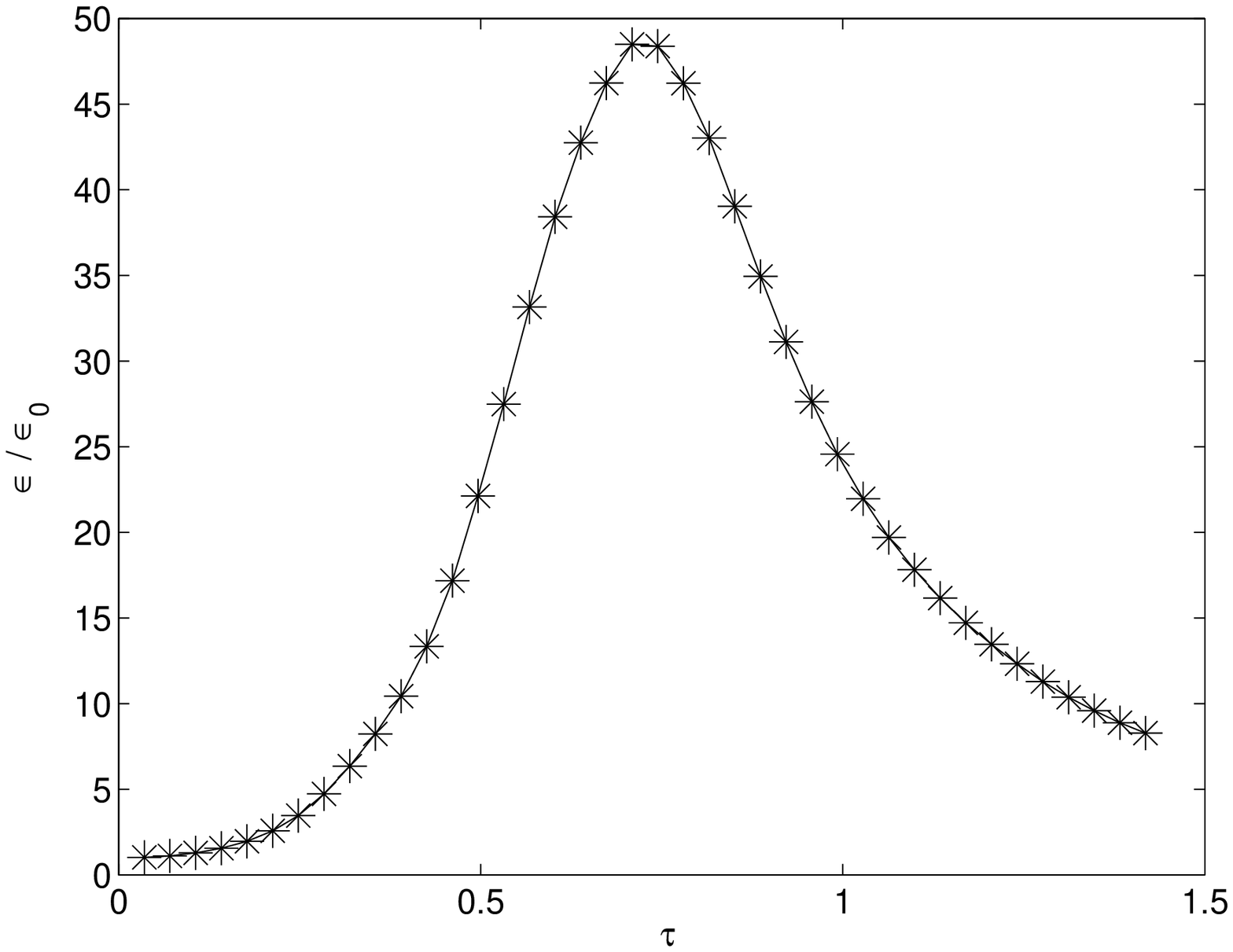}
\caption{\label{prelim}(a) Log-log plot of the temporal evolution of the 
scaled kinetic energy spectrum $k^{5/3}E(k,\tau)$ as a function of 
the wavenumber $k$ at temporal separations of $\tau=0.24$, with 
$E(k,t_0)\sim k^4e^{-2k^2}$. The 
plot with open circles is calculated at cascade completion, at dimensionless 
time $\tau=\tau_c=0.71$.\\
(b) Plot of the normalized kinetic energy $E(\tau)/E_0$ as a function of the 
dimensionless time $\tau$, with $E(k,t_0)\sim k^4e^{-2k^2}$.\\(c) Plot of the 
normalized kinetic 
energy-dissipation rate $\epsilon(\tau)/\epsilon_0$ as a function of the 
dimensionless time $\tau$, with $E(k,t_0)\sim k^4e^{-2k^2}$.} 
\end{figure}

We use a pseudospectral method\cite{Dhar} to solve Eqs. (\ref{nse}) 
numerically, in a cubical box of side $2\pi$ with periodic boundary conditions 
and $192^3$ 
Fourier modes. We do not address issues pertaining to the scaling of 
higher-order structure functions of velocity differences or investigate 
dissipation-scale properties, and believe that our spectral resolution is 
adequate for the types of studies that we have carried out. For the temporal 
evolution, we use an Adams-Bashforth scheme 
(step size $\delta t=10^{-3}$) with double-precision arithmetic and set 
$\nu=10^{-5}$. We include a hyperviscous term of the form 
$\nu_h\nabla^4{\bf v}$ in Eqs. (\ref{nse}), with $\nu_h=10^{-6}$. We show 
explicitly in Section 
II(C), that our results are unaffected by inclusion of hyperviscosity. The 
initial velocity field is taken to be 
${\bf v}({\bf k},t_0)\sim k^2e^{-k^2}e^{i\theta_{\bf k}}$, with 
$\theta_{\bf k}$ random variables distributed uniformly between $0$ and 
$2\pi$. This corresponds to an 
initial energy spectrum $E(k,t_0)\sim k^4e^{-2k^2}$ (with $E(k,t)\equiv|{\bf v}
({\bf k},t)|^2$, the one-dimensional spectrum), which is a convenient choice 
that develops a cascade to large wavenumbers (see below). 
We measure time in units of the initial large-eddy turnover time 
$\tau_0\equiv2\pi/v_{rms}^0$ (here  $\tau_0$ equals $4.23$), 
$v^0_{rms}\equiv[\langle\sum_{{\bf k}}|{\bf v}({\bf k},t_0)|^2\rangle]^{1/2}$ 
is the root-mean-square value of the initial velocity, with the dimensionless 
time $\tau\equiv t/\tau_0$ ($t$ is the product of the number of steps and 
$\delta t$). We define $Re_0\equiv2\pi v^0_{rms}/\nu$ to be the value of the 
initial `box-size' Reynolds number (here $Re_0$ equals $932965$). Our 
results are obtained for times $t_0\le t<<t_*$, where 
$t_*$ is the time at which the (growing) integral scale 
$L(t)\equiv\langle(\sum_{\bf k}|{\bf v}({\bf k},t)|^2/k)/\sum_{\bf k}
|{\bf v}({\bf k},t)|^2\rangle$ becomes of the order of the linear size of the 
simulation box. For times $t\gtrsim t_*$, finite-size effects which might 
well be non-universal, modify the numerical results, and are not considered 
here.\\
In Fig. \ref{prelim}, we show some preliminary results that serve as a 
check of our numerical method and parameter values (which were chosen 
to ensure linear stability of the numerical scheme). 
Figure \ref{prelim}(a) shows on a log-log plot, the scaled kinetic energy 
spectrum $k^{5/3}E(k,\tau)$ as a function of the 
wavenumber $k$. On starting with the spectrum specified above, 
a cascade of energy is seen to large wavenumbers. The plots are 
equispaced in time with a temporal separation of $\tau=0.24$. The plot with 
open circles is calculated at cascade completion at dimensionless time 
$\tau=\tau_c=0.71$, 
and shows a wavenumber range (for $1\lesssim k\lesssim10$) that exhibits the 
well-known $-5/3$ power-law\cite{Yamamoto}. Upon 
cascade completion, the shape of the energy spectrum does not change 
appreciably (except at large wavenumbers where it falls), but the kinetic 
energy decays. In Fig. \ref{prelim}(b), we plot the normalized kinetic 
energy $E(\tau)/E_0$ as a 
function of the dimensionless time $\tau$, which is found to decay 
monotonically\cite{Yamamoto}. In 
Fig. \ref{prelim}(c), we plot the normalized kinetic energy-dissipation rate 
$\epsilon(\tau)/\epsilon_0$ [$\epsilon(t)\equiv\sum_{\bf k}k^2|{\bf v}
({\bf k},t)|^2$] 
as a function of the dimensionless time $\tau$. The kinetic energy-dissipation 
rate peaks\cite{Yamamoto,Hosokawa} at $\tau=\tau_c$, corresponding to cascade 
completion in the 
energy spectrum, and decreases thereafter. The turbulence may be 
considered as `fully developed' at $\tau=\tau_c$ and our spatial results 
(see below) will be calculated at this instant of time.
\subsection{Vorticity}
\begin{figure}
\includegraphics[height=2.0in]{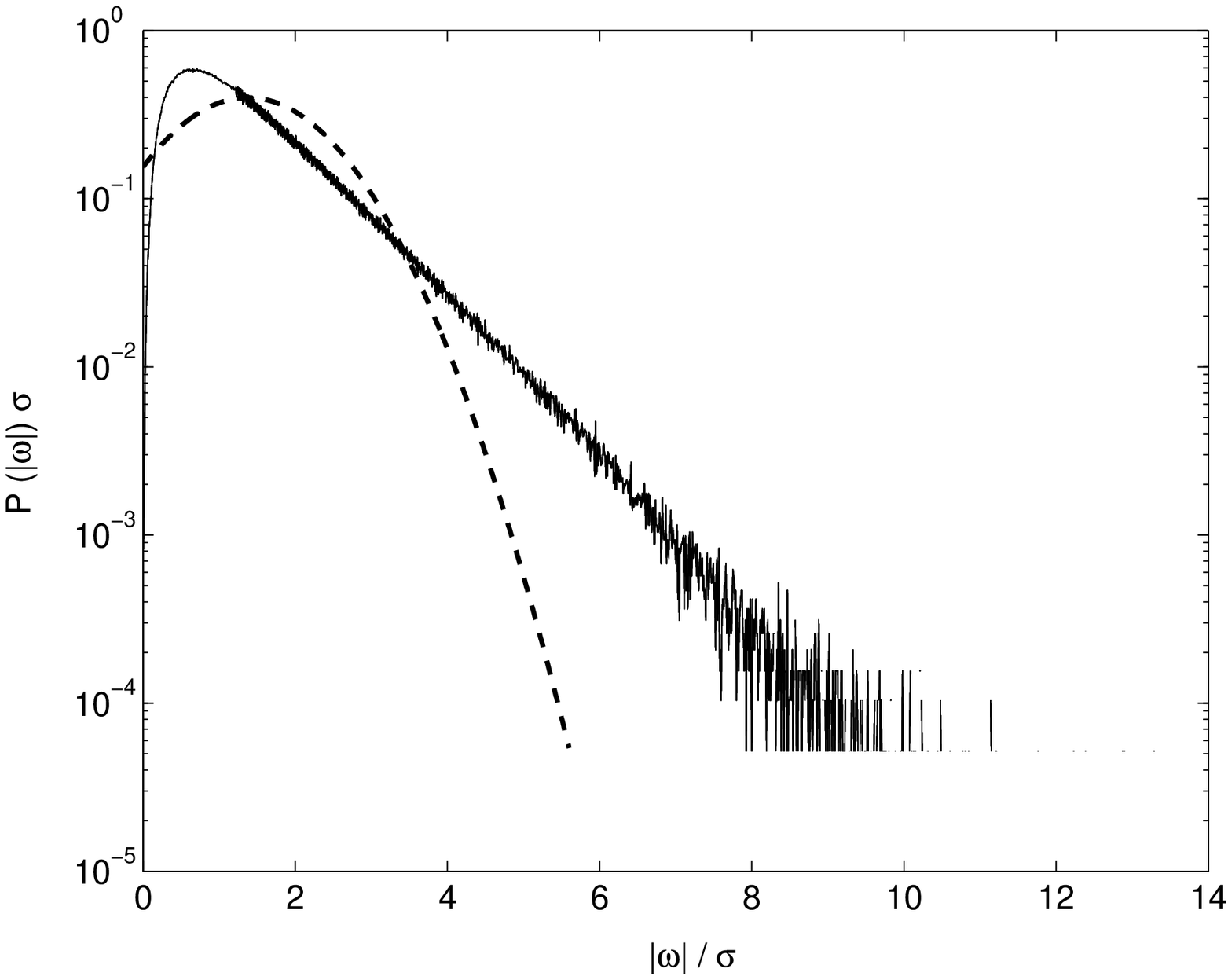}
\includegraphics[height=2.0in]{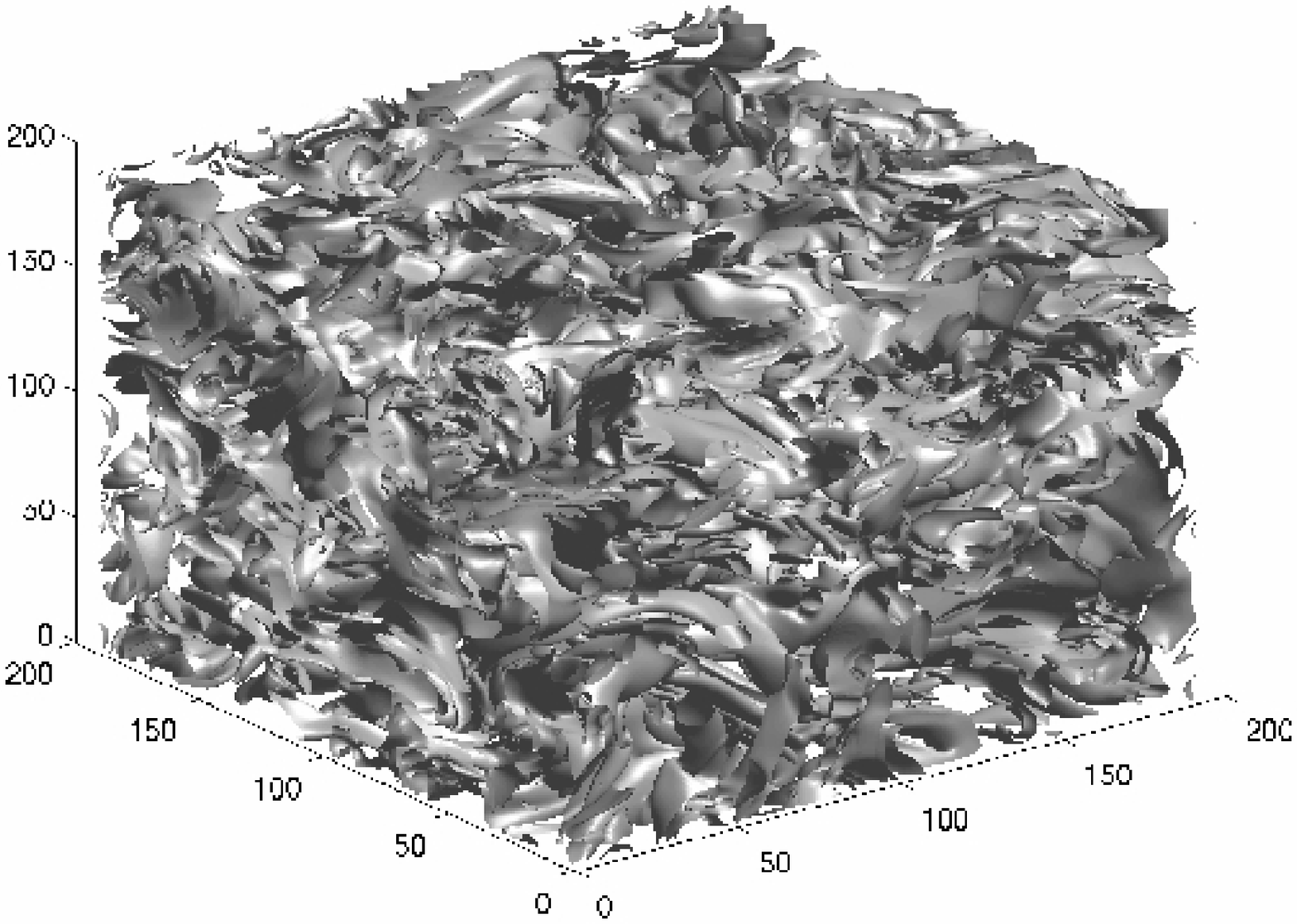}
\caption{\label{vort}(a) Semilog plot of the normalized 
probability distribution $P(|\omega|)$ of $|\omega|$ ($\omega$ is the 
vorticity) at cascade completion, with $E(k,t_0)\sim k^4e^{-2k^2}$. The 
dashed-line plot is a normalized Gaussian distribution for comparison.\\
(b) Plot of iso-$|\omega|$ surfaces for 
the isovalue $|\omega|=\langle|\omega|\rangle$ at cascade completion, with 
$E(k,t_0)\sim k^4e^{-2k^2}$.}
\end{figure}

\begin{figure}
\includegraphics[height=2.0in]{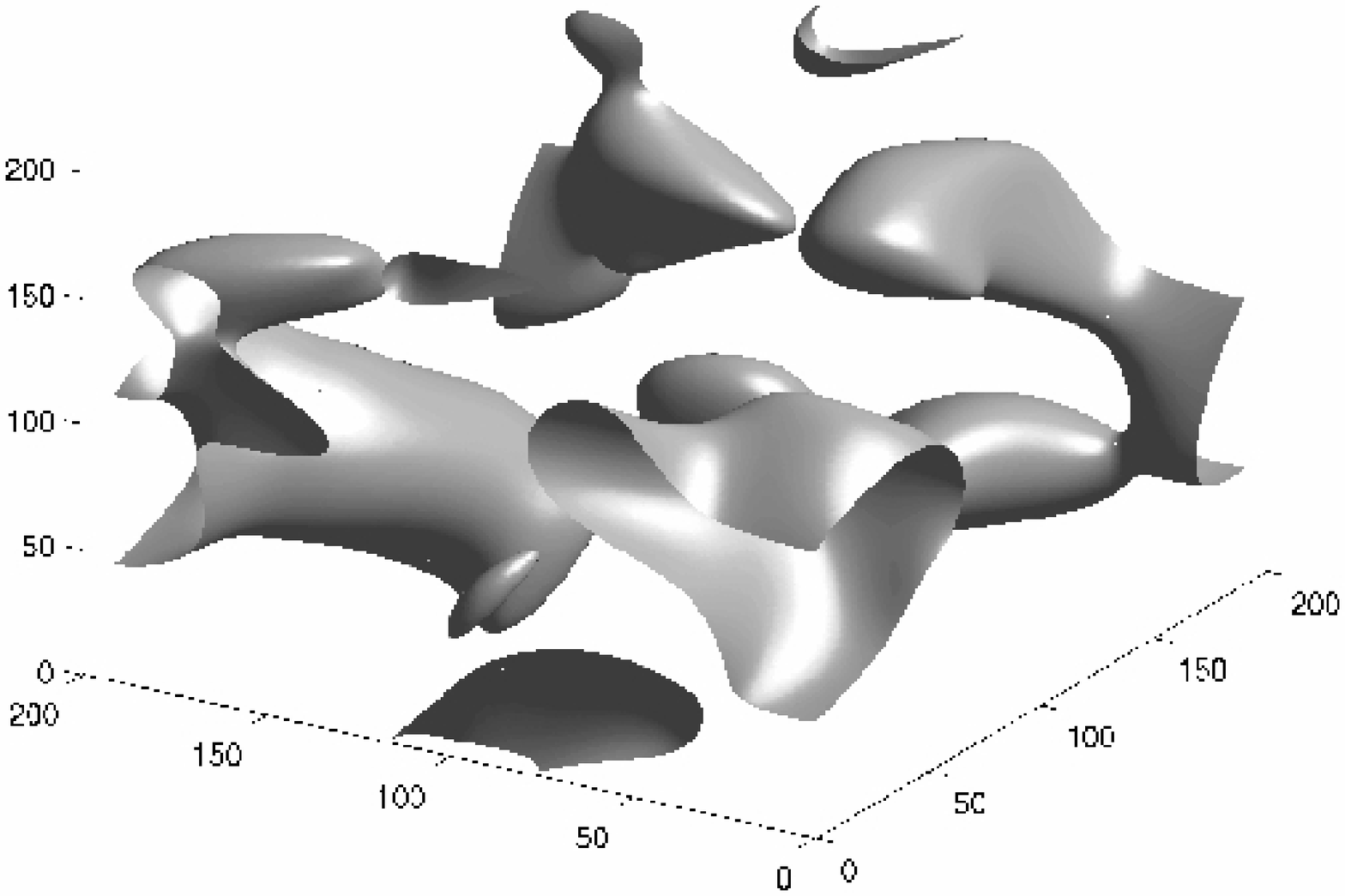}
\includegraphics[height=2.0in]{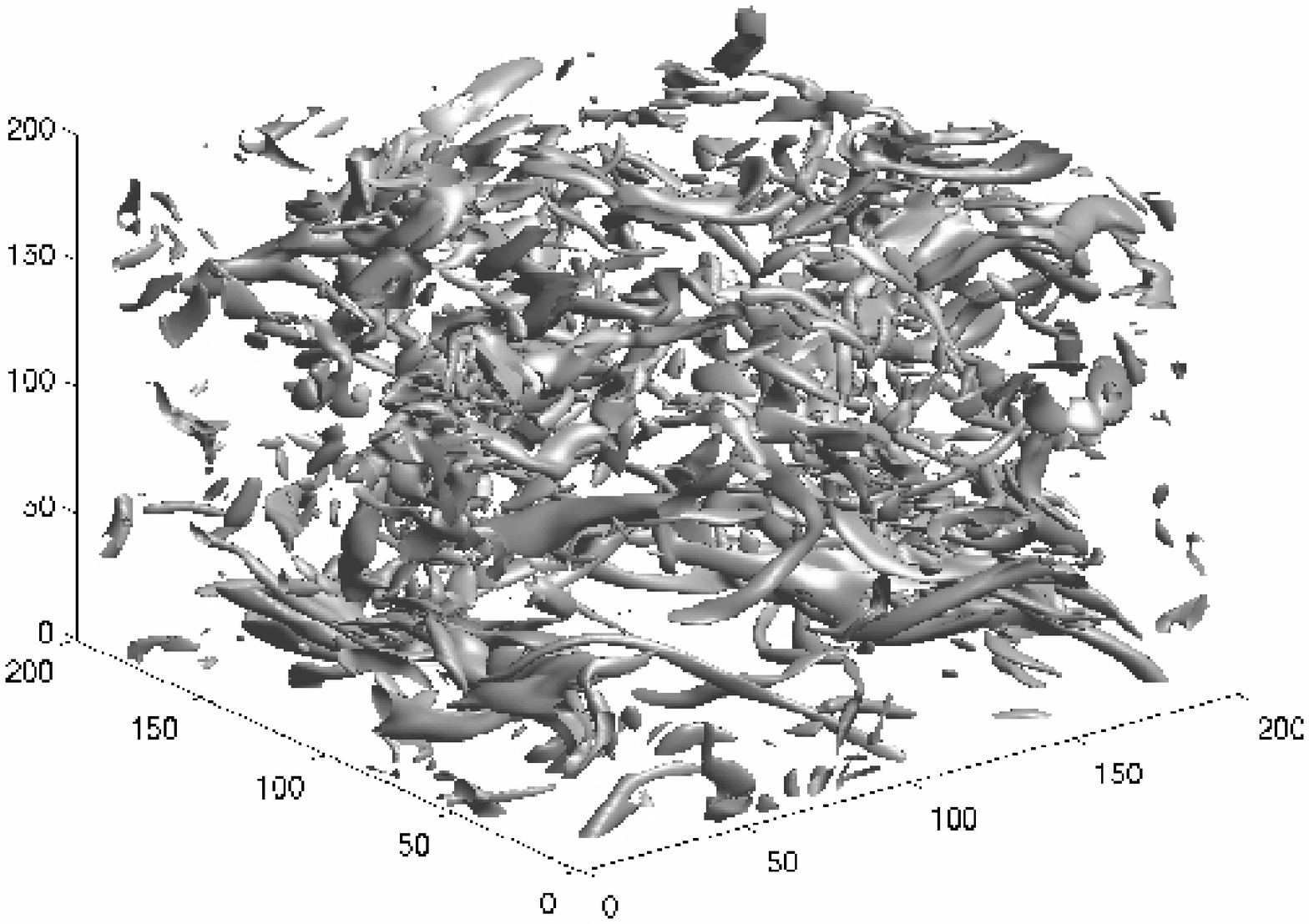}
\caption{\label{vortmax}(a) Plot of iso-$|\omega|$ surfaces for 
the isovalue $|\omega|=\langle|\omega|\rangle+2\sigma$ ($\sigma$ is 
the standard deviation) at dimensionless time 
$\tau<<\tau_c$, with $E(k,t_0)\sim k^4e^{-2k^2}$.\\
(b) Plot of iso-$|\omega|$ surfaces for 
the isovalue $|\omega|=\langle|\omega|\rangle+2\sigma$ at cascade 
completion, with $E(k,t_0)\sim k^4e^{-2k^2}$.} 
\end{figure}

The dynamics of vortex structures in a turbulent flow-field is governed by the 
response of the vorticity vector to the strain-rate tensor. In 
Fig. \ref{vort}(a), we plot the normalized probability 
distribution $P(|\omega|)$ of $|{\bf\omega}|$ at cascade completion. The 
distribution is found to peak at 
$|\omega|=1.5\sigma$ ($\sigma$ denotes the standard deviation) and has an 
exponential tail for $|\omega|>1.5\sigma$ with a fit 
$P(|\omega|)\sim e^{-\beta|\omega|^\alpha}$, $\beta=0.05\pm0.01$, 
$\alpha=1.00\pm0.01$ (error-bars from a least-squares fit). In 
Fig. \ref{vort}(b), we plot iso-$|{\bf\omega}|$
surfaces for the isovalue $|\omega|=\langle|\omega|\rangle$ (angular brackets 
denote a volume average) at cascade completion, which appear to be crumpled 
sheet-like structures (found throughout the isovalue range 
$[\langle|\omega|\rangle-\sigma,\langle|\omega|\rangle+\sigma]$). At early 
times $\tau<<\tau_c$, regions of intense vorticity (with the isovalue 
$|\omega|=\langle|\omega|\rangle+2\sigma$) were found to be 
sheet-like (see Fig. \ref{vortmax}(a)). It is plausible that a 
small volume fraction\cite{smallfn} of such structures, subject to shear and 
stretching, `roll-up' to form the filaments visualised in 
Fig. \ref{vortmax}(b) at cascade completion.  The filaments are 
found to be of diameter of the order of the grid spacing, and have a contour 
length that occasionally 
extends nearly to the linear size of the simulation box\cite{hosfn}. We 
choose to quote 
dimensions of the structures relative to the (fixed) box-size and the 
grid-spacing, since both the dissipative and the integral length scales vary in 
time, in decaying turbulence. Iso-$|\omega|$ 
surfaces in the range $|\omega|<(\langle|\omega|\rangle-\sigma)$ were not 
found to exhibit any particular structure at cascade completion. 
The above results are in accordance with earlier studies of statistically 
steady\cite{Douady,Siggia,Vincent,Jimenez} and decaying\cite{Yamamoto,Hosokawa} 
turbulence.
\subsection{Strain-Rate Tensor}
\begin{figure}
\includegraphics[height=2.0in]{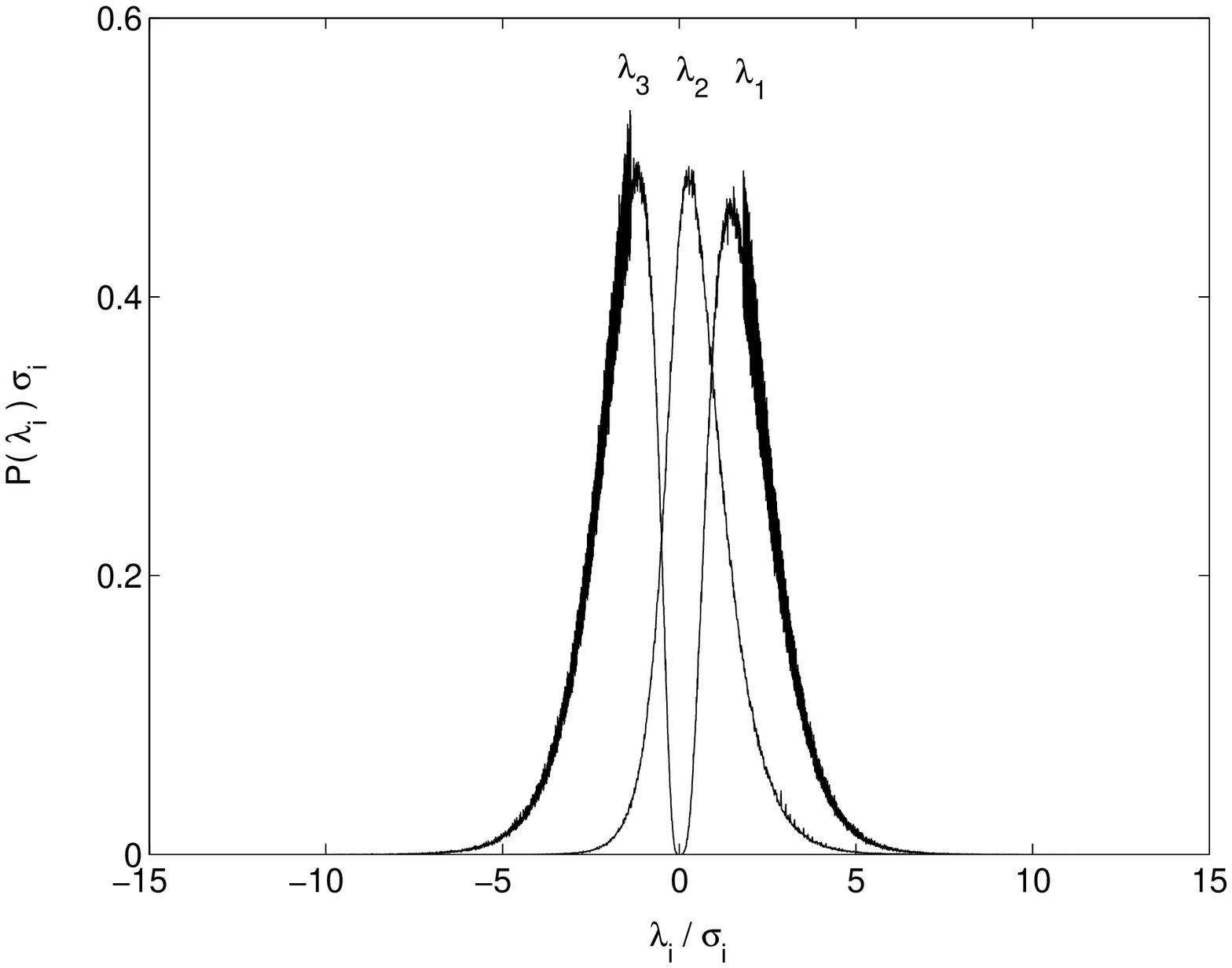}
\includegraphics[height=2.0in]{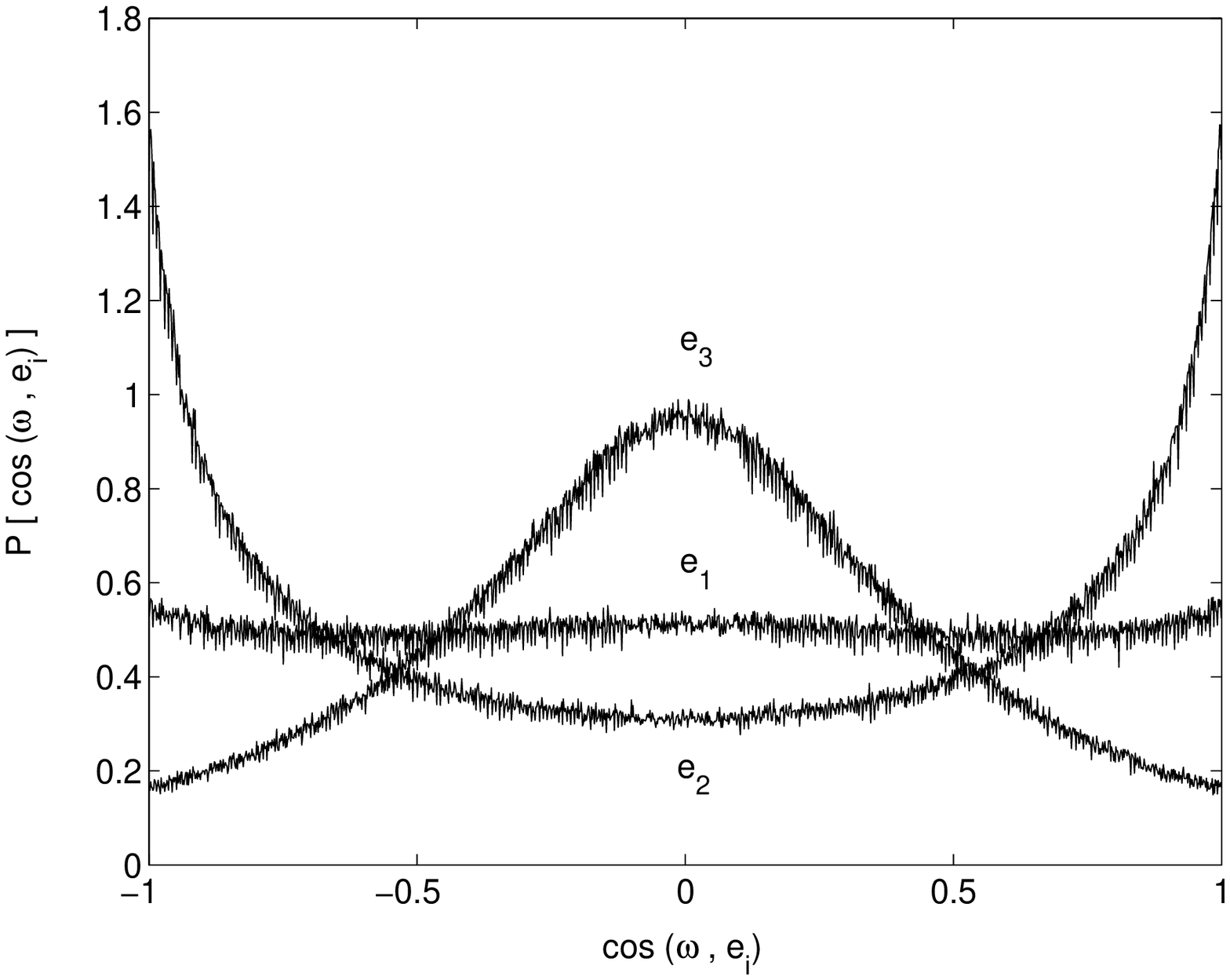}
\includegraphics[height=2.0in]{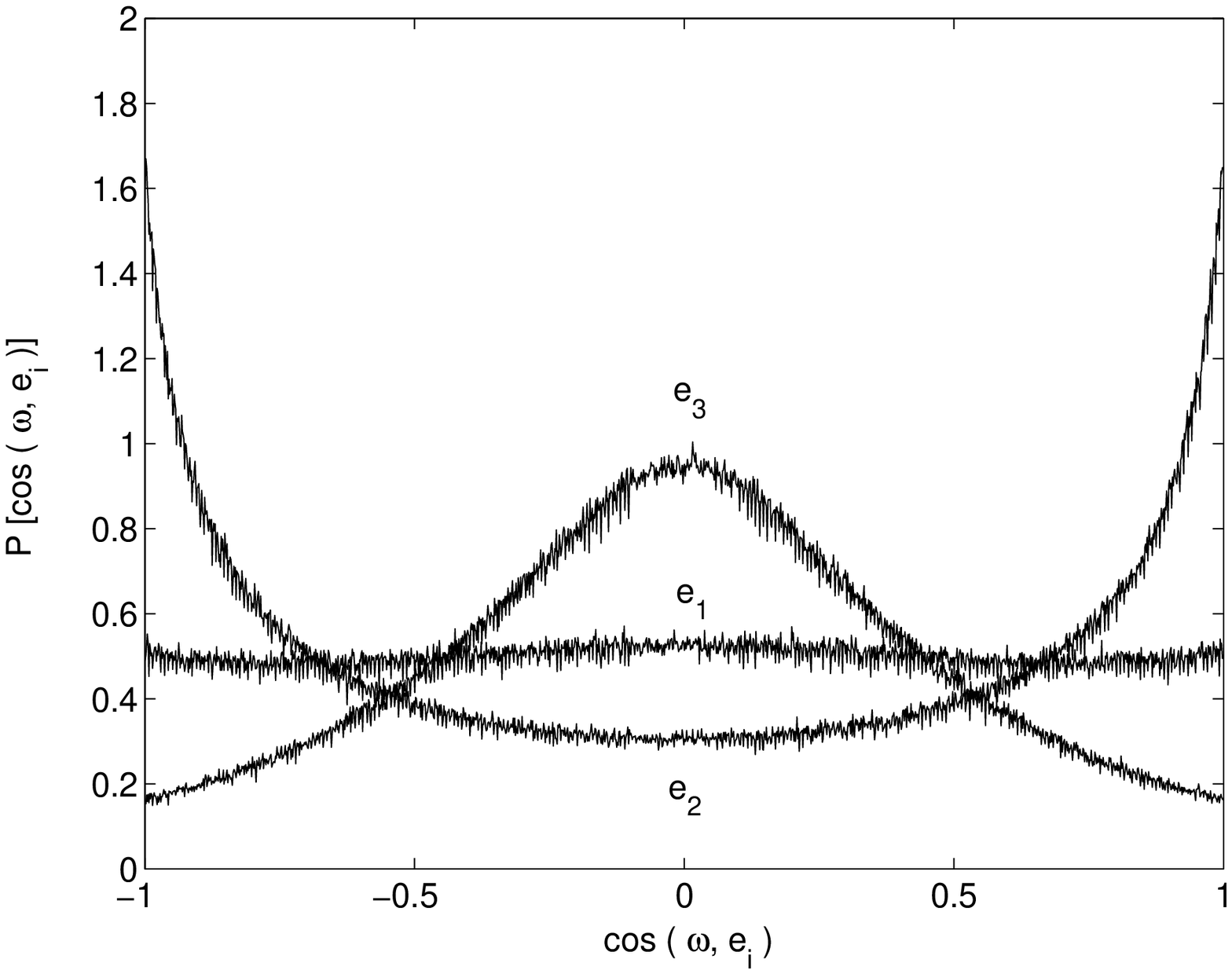}
\caption{\label{eig}(a) Plot of the normalized probability distribution
$P(\lambda_i)$ of the eigenvalues 
$\lambda_i$ of the strain-rate tensor $S_{ij}$, $i,j=1,2,3$, at cascade 
completion, with $E(k,t_0)\sim k^4e^{-2k^2}$.\\
(b) Plot of the normalized probability distribution of cosine of 
the angle between $\omega$ and the 
eigenvectors $e_i$ of $S_{ij}$ at cascade completion, with 
$E(k,t_0)\sim k^4e^{-2k^2}$.\\
(c) Plot of the normalized probability distribution of cosine of the 
angle between $\omega$ and the eigenvectors $e_i$ at cascade completion, 
with $E(k,t_0)\sim k^4e^{-2k^2}$ and without hyperviscosity.}
\end{figure}

At each grid point, we compute the eigenvalues $\lambda_1$, $\lambda_2$ and 
$\lambda_3$ (with the convention $\lambda_1\ge\lambda_2\ge\lambda_3$) of 
the strain-rate tensor $S_{ij}$, and the 
corresponding orthonormal eigenvectors $e_1$, $e_2$, and $e_3$. In 
a constant-density 
flow, incompressibility requires that $\sum_i\lambda_i=0$, with 
$\lambda_1>0$, $\lambda_3<0$, the sign of the intermediate eigenvalue 
$\lambda_2$ being indeterminate. In Fig. \ref{eig}(a), we plot the normalized 
probability 
distribution $P(\lambda_i)$ of the eigenvalues $\lambda_i$ at 
cascade completion. We find that $\lambda_2$ has a positive mean. The 
statistically preferred ratio of the mean strain-rates 
$\langle\lambda_1\rangle:\langle\lambda_2\rangle:\langle\lambda_3\rangle$ was 
found to equal $4.9:1:-5.9$ at cascade completion\cite{modefn}.

In Fig. \ref{eig}(b), we plot the normalized probability distribution 
$P[\cos(\omega,e_i)]$ of cosine of the angle between $\omega$ and the 
eigenvectors $e_i$, at cascade completion. 
We observe that $\omega$ is preferentially aligned (or antialigned) with 
eigenvector $e_2$ (a peak-to-valley ratio $\approx5.1$ was found) 
corresponding to eigenvalue $\lambda_2$ of {\it minimum} 
relative mean strain-rate (disregarding the sign in the mean ratio, which 
merely indicates an extensional or compressional strain on the fluid element). 
The vorticity is also found to 
be preferentially perpendicular to eigenvector $e_3$ corresponding to the 
principal compressive eigenvalue $\lambda_3$, while no specific angular 
relationship is observed relative to eigenvector $e_1$. We could 
confirm an earlier result\cite{Vincent} that the preferential alignment of 
$\omega$ with $e_2$ is already pronounced {\it during} the 
cascade process, much prior to the appearance of distinct filaments in 
regions of intense vorticity. In Fig. \ref{eig}(c), we plot 
$P[\cos(\omega,e_i)]$ at cascade completion, for a run with the same parameter 
values as in Section II(A), but without hyperviscosity (here $\nu=10^{-4}$). 
We find that the inclusion of hyperviscosity introduces no qualitative 
changes (a peak-to-valley ratio of $5.3$ was found), and we believe that 
effects on structural properties are minimal. The positive mean of 
$\lambda_2$, the ratio of the mean strain-rates, and the 
preferential alignment of 
$\omega$ with $e_2$ is consistent with corresponding observations in 
statistically steady\cite{Vincent} and decaying\cite{Ashurst} turbulence.
\subsection{Mean Enstrophy Production Rate}
\begin{figure}
\includegraphics[height=2.0in]{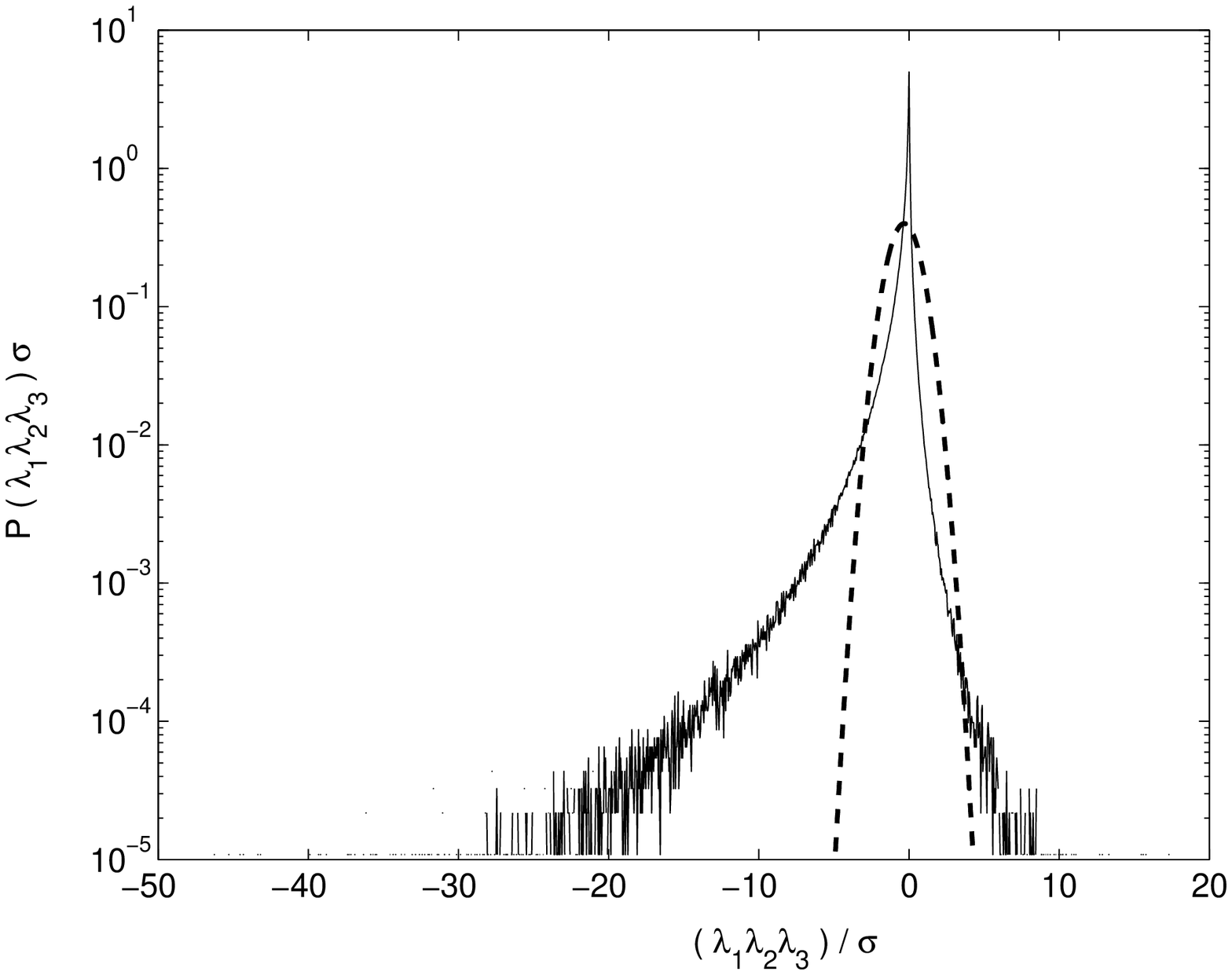}
\includegraphics[height=2.0in]{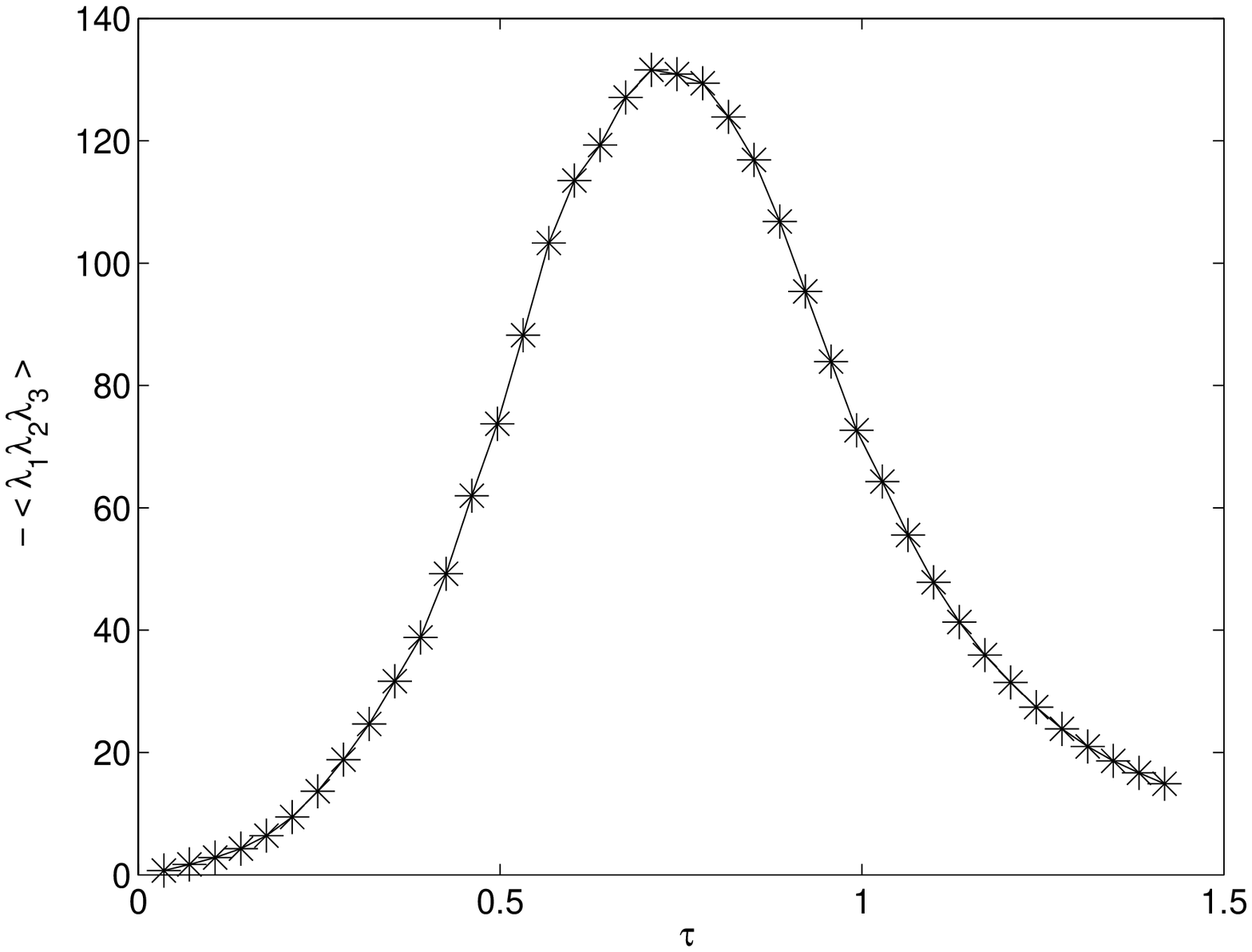}
\includegraphics[height=2.0in]{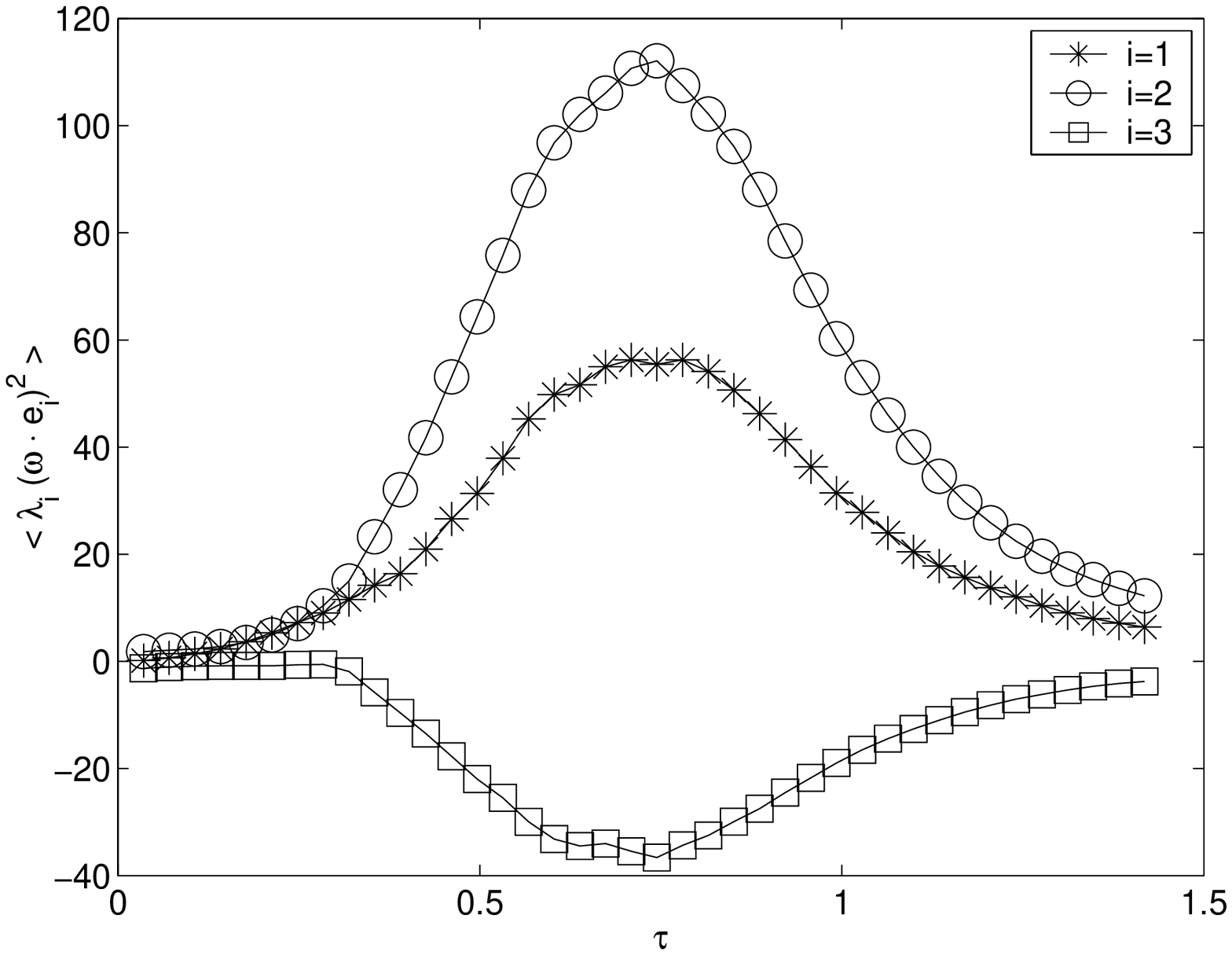}
\caption{\label{betchov}(a) Semilog plot of the normalized 
probability distribution 
$P(\lambda_1\lambda_2\lambda_3)$ of $\lambda_1\lambda_2\lambda_3$ 
at cascade completion, with $E(k,t_0)\sim k^4e^{-2k^2}$. The 
dashed-line plot is a normalized Gaussian distribution for comparison.\\
(b) Plot of $-\langle\lambda_1\lambda_2\lambda_3\rangle$ (see Eq. 
(\ref{betchoveqn})) as a function of the dimensionless time $\tau$, with 
$E(k,t_0)\sim k^4e^{-2k^2}$.\\
(c) Plot of the rates of production of the 
mean enstrophy $\langle\lambda_i(\omega\cdot e_i)^2\rangle$ along the 
eigenvectors $e_i$ of $S_{ij}$, as a function of 
the dimensionless time $\tau$, with $E(k,t_0)\sim k^4e^{-2k^2}$.}
\end{figure}

\begin{figure}
\includegraphics[height=2.0in]{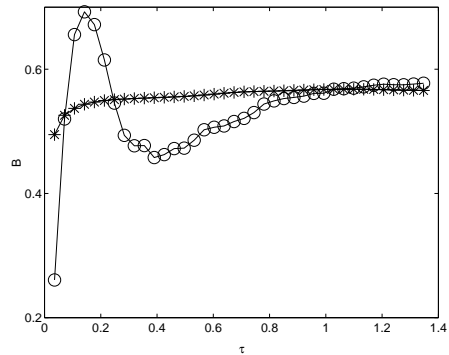}
\caption{\label{betineq}Plot of Betchov's inequality 
$B\equiv 3\sqrt{6}|\langle\lambda_1\lambda_2\lambda_3\rangle|/
\langle(\lambda_1^2+\lambda_2^2+\lambda_3^2)^{3/2}\rangle\le 1$ as a function 
of the dimensionless time $\tau$, with $E(k,t_0)\sim k^4e^{-2k^2}$ 
(open circles) and $E(k,t_0)\sim k$ (asterisks).}
\end{figure}

In a classical work\cite{BetchovJFM}, Betchov derived 
an equation for the rate of  
production of the mean enstrophy $\langle\omega^2\rangle$\cite{Redfn} in 
unforced, incompressible, homogeneous, and isotropic turbulence, viz.
\begin{eqnarray}
&&\frac{\partial\langle{\omega^2}\rangle}{\partial{t}}=
-\langle\lambda_1\lambda_2\lambda_3\rangle-\frac{\nu}{4}\left<\left
(\frac{\partial^2v_i}{\partial x_j\partial x_k}\right)^2\right>,
\label{betchoveqn}
\end{eqnarray}
where the angular brackets denote a volume average. 
From Eq. (\ref{betchoveqn}), it is evident that production of 
$\langle\omega^2\rangle$ requires a predominantly negative 
value of $\lambda_1\lambda_2\lambda_3$ in the 
fluid. In Fig. \ref{betchov}(a), we plot the normalized probability 
distribution $P(\lambda_1\lambda_2\lambda_3)$ of $\lambda_1\lambda_2\lambda_3$ 
at cascade completion and find that the distribution has a negative mean (with 
a skewness equal to $-7.83$), a result that has already been noted in an 
early numerical study\cite{BetchovPOF}. If $\lambda_1\lambda_2\lambda_3$ is 
positive, we have one positive and two negative eigenvalues: a vortex element 
is stretched along one direction and squashed along the other two, forming a 
slender filamentary structure. If $\lambda_1\lambda_2\lambda_3$ is negative, 
we have one negative and two positive eigenvalues: a vortex element is 
stretched along two directions and squashed along the third, forming a 
sheet-like structure. In homogeneous turbulence 
$\langle\lambda_1\lambda_2\lambda_3\rangle$ is found to be 
negative\cite{BetchovJFM,Townsend}, 
indicating a predominance of sheet-like structures, which accords 
with Fig. \ref{vort}(b). In Fig. \ref{betchov}(b), we plot 
$-\langle\lambda_1\lambda_2\lambda_3\rangle$ as a function of the 
dimensionless time $\tau$. We find that the magnitude of 
$-\langle\lambda_1\lambda_2\lambda_3\rangle$ peaks at cascade completion 
($\tau=0.71$), which is 
consistent with the dimensionless time at which the kinetic energy-dissipation 
rate is a maximum (cf. Fig. \ref{prelim}(c)). In homogeneous turbulence, 
the term $-\langle\lambda_1\lambda_2\lambda_3\rangle$ can be 
shown\cite{BetchovJFM} to equal 
the expression $\langle\omega_i\omega_jS_{ij}\rangle=\sum_i\langle\lambda_i
(\omega\cdot e_i)^2\rangle$. 
In Fig. \ref{betchov}(c), we plot $\langle\lambda_i(\omega\cdot e_i)^2\rangle$, 
the rates of production of the mean enstrophy along 
eigenvectors $e_i$, and 
find that the production rate is largest along eigenvector $e_2$ corresponding 
to eigenvalue $\lambda_2$ of {\it minimum} relative mean strain-rate 
(see above), a remarkable result that appears to be new.\\
From Fig. \ref{betchov}(b) we observe that the mean enstrophy production rate 
$-\langle\lambda_1\lambda_2\lambda_3\rangle$ has an upper bound. In the same 
work\cite{BetchovJFM}, Betchov invoked formal mathematical 
inequalities and incompressibility to derive the upper bound
\begin{eqnarray}
|\langle\lambda_1\lambda_2\lambda_3\rangle|\le \frac{1}{3\sqrt{6}}
\langle(\lambda_1^2+\lambda_2^2+\lambda_3^2)^{3/2}\rangle.
\label{betchovineq}
\end{eqnarray}
The upper bound is the largest rate of production of mean enstrophy 
compatible with the requirements of isotropy, homogeneity, and 
incompressibility. In Fig. \ref{betineq}, we plot inequality 
(\ref{betchovineq})\cite{Betchovfn} as a function of the dimensionless 
time $\tau$ and find that the inequality is satisfied to a greater degree 
{\it during} the cascade process, and for dimensionless times 
$\tau\gtrsim\tau_c$ asymptotes to the 
value $\approx0.57$. Dynamical effects suggest an inequality that is 
stronger than the one due to Betchov.
\section{Power-Law Spectrum}
\subsection{Numerical Method}
\begin{figure}
\includegraphics[height=2.0in]{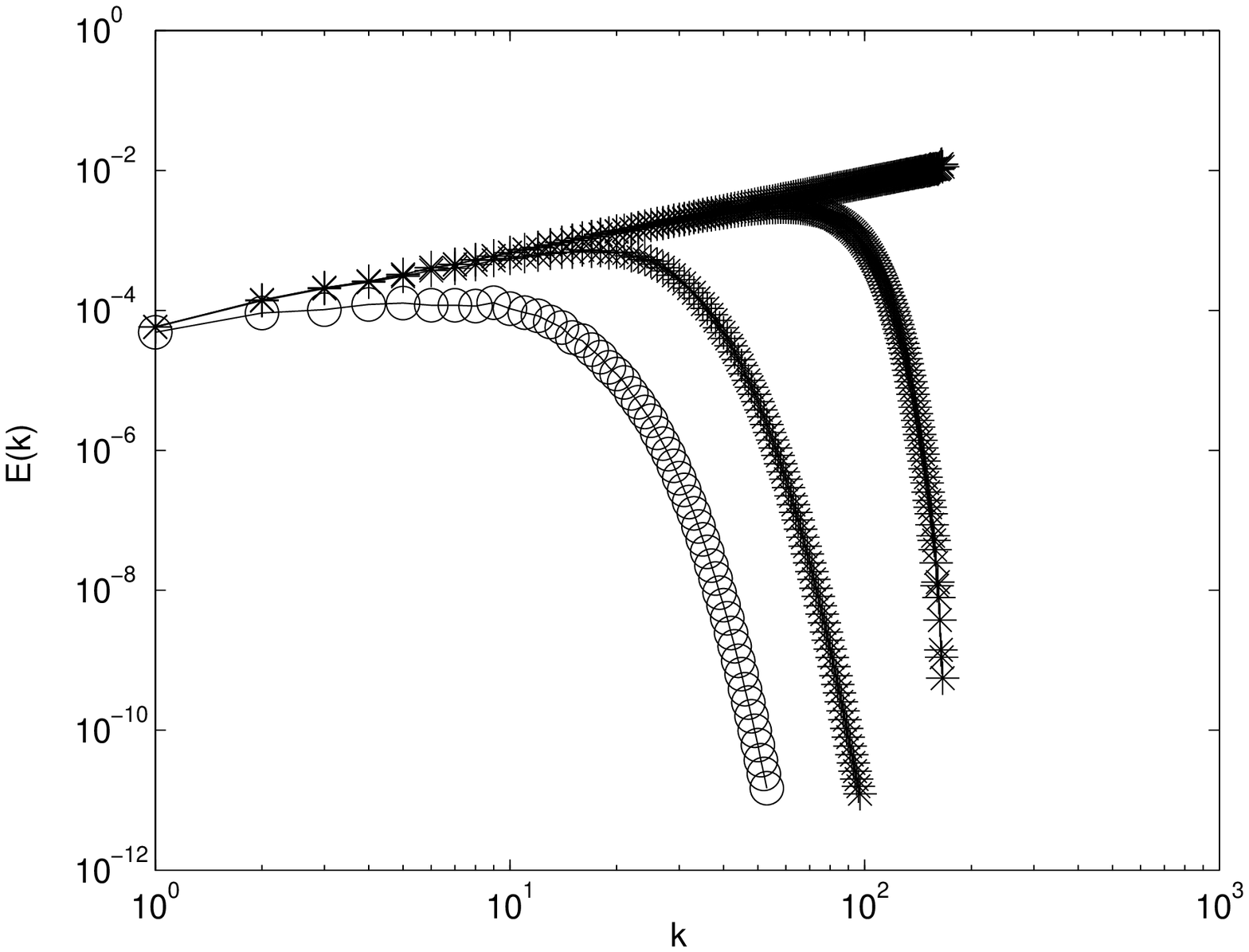}
\includegraphics[height=2.0in]{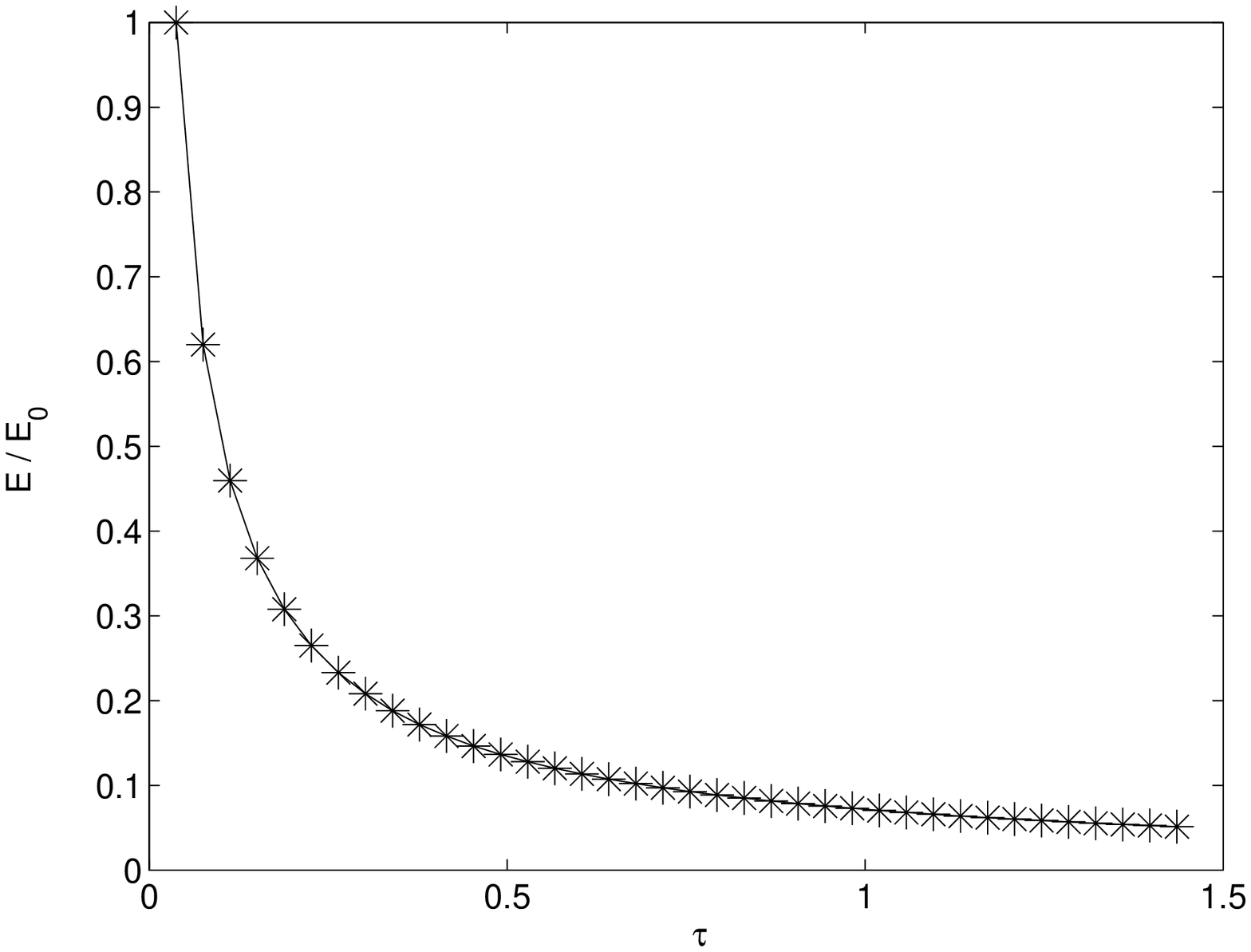}
\includegraphics[height=2.0in]{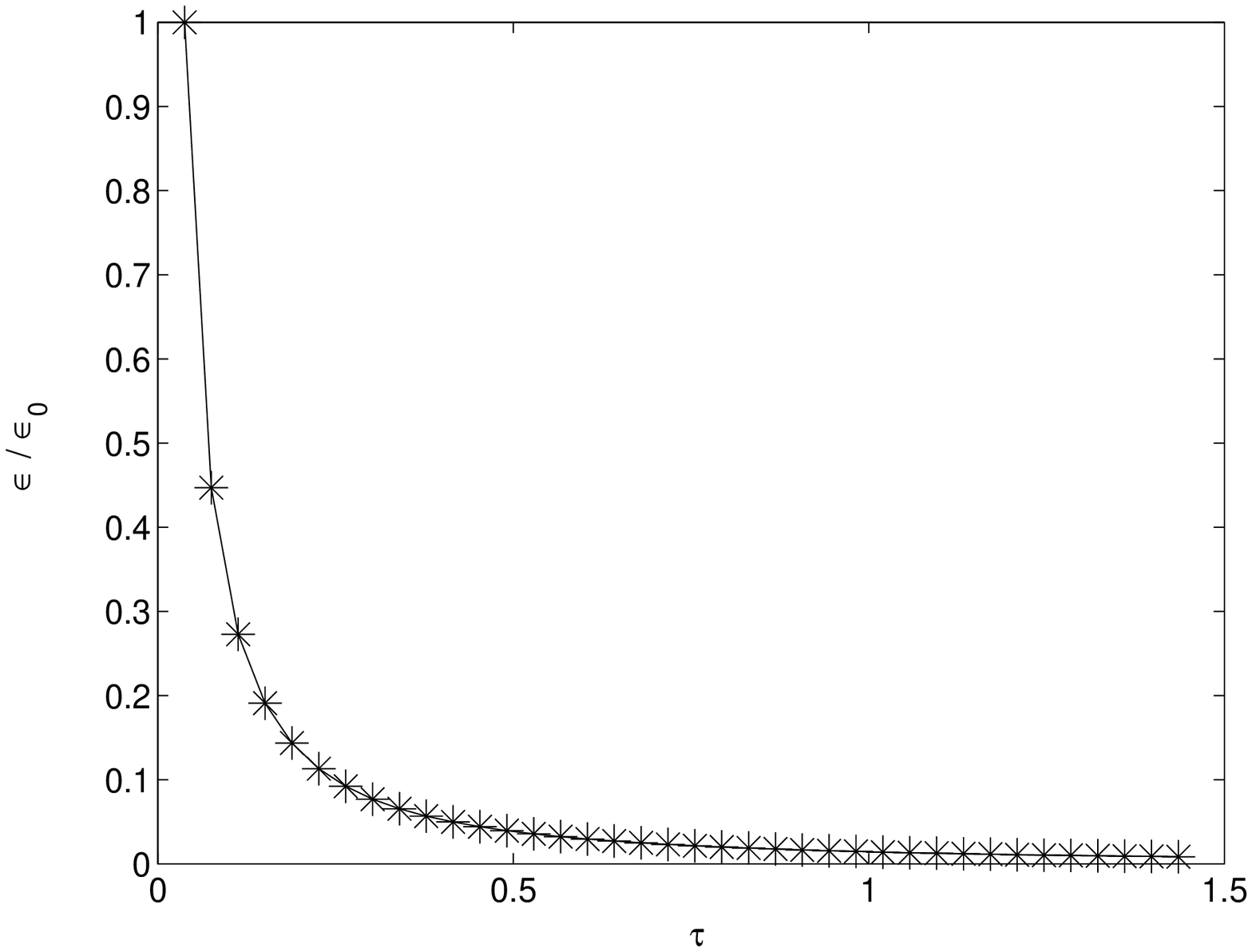}
\caption{\label{prelimq1}(a) Log-log plot of the temporal evolution of the 
kinetic energy spectrum $E(k,\tau)$ as a function of the wavenumber $k$ at temporal seperations of $\tau=0.24$, with $E(k,t_0)\sim k$. The 
plot with open circles is calculated at dimensionless time 
$\tau=\tau_c=0.71$.\\ 
(b) Plot of the temporal evolution of the 
normalized kinetic energy $E(\tau)/E_0$ as a function of the dimensionless 
time $\tau$, with $E(k,t_0)\sim k$. \\(c) Plot of the normalized kinetic 
energy-dissipation rate 
$\epsilon(\tau)/\epsilon_0$ as a function of the dimensionless time $\tau$, 
with $E(k,t_0)\sim k$.} 
\end{figure}

We use the numerical scheme, spectral resolution, step size, viscosities, and 
boundary conditions as specified in Section II(A). The initial velocity 
field is taken 
to be ${\bf v}({\bf k},t_0)\sim k^{1/2}e^{i\phi_{\bf k}}$, with 
$\phi_{\bf k}$ random variables distributed uniformly between $0$ and 
$2\pi$, which corresponds to the 
initial energy spectrum $E(k,t_0)\sim k$ (with 
$E(k,t)\equiv|{\bf v}({\bf k},t)|^2$). Here, the dimensionless time $\tau_0$ 
equals $13.9$ and the initial `box-size' Reynolds number $Re_0$ equals 
$350181$. 
Our results are obtained for times $t_0\le t<<t_*$, and we choose to calculate 
our spatial 
results at dimensionless time $\tau=\tau_c=0.71$\cite{Refn} which is equal to 
the dimensionless 
time at which the kinetic energy-dissipation rate peaks, on starting with a 
`cascade-type' spectrum as shown in Fig. \ref{prelim}(c).\\
In Fig. \ref{prelimq1}(a), we show on a log-log 
plot, the kinetic energy spectrum $E(k,\tau)$ as a function of the 
wavenumber $k$. The plots are equispaced in time with a temporal separation of 
$\tau=0.24$ and the plot with open circles is calculated at $\tau=\tau_c$. 
The spectrum does not cascade to large wavenumbers or exhibit 
a wavenumber range with a $-5/3$ power-law at any stage of temporal 
evolution. In Figs. \ref{prelimq1}(b) and (c), 
we plot the normalized kinetic energy $E(\tau)/E_0$ and the normalized 
kinetic energy-dissipation rate $\epsilon(\tau)/\epsilon_0$, as 
a function of the dimensionless time $\tau$ (plotted on a linear scale in 
order to compare with Figs. \ref{prelim}(b) and (c)). The kinetic energy 
$E(\tau)$ is 
found to decay as a power-law (on a log-log plot) with an exponent equal 
to $-0.91\pm0.04$, with error-bars from a least-squares fit. The 
exponent is theoretically predicted\cite{Ditlevsen} to 
equal $-1$, and 
we believe the discrepancy is due to the low spectral resolution of our DNS. 
The normalized kinetic energy-dissipation rate $\epsilon(\tau)/\epsilon_0$ 
does not exhibit a peak (cf. Fig. \ref{prelim}(c)) and decays monotonically.
\subsection{Vorticity}
\begin{figure}
\includegraphics[height=2.0in]{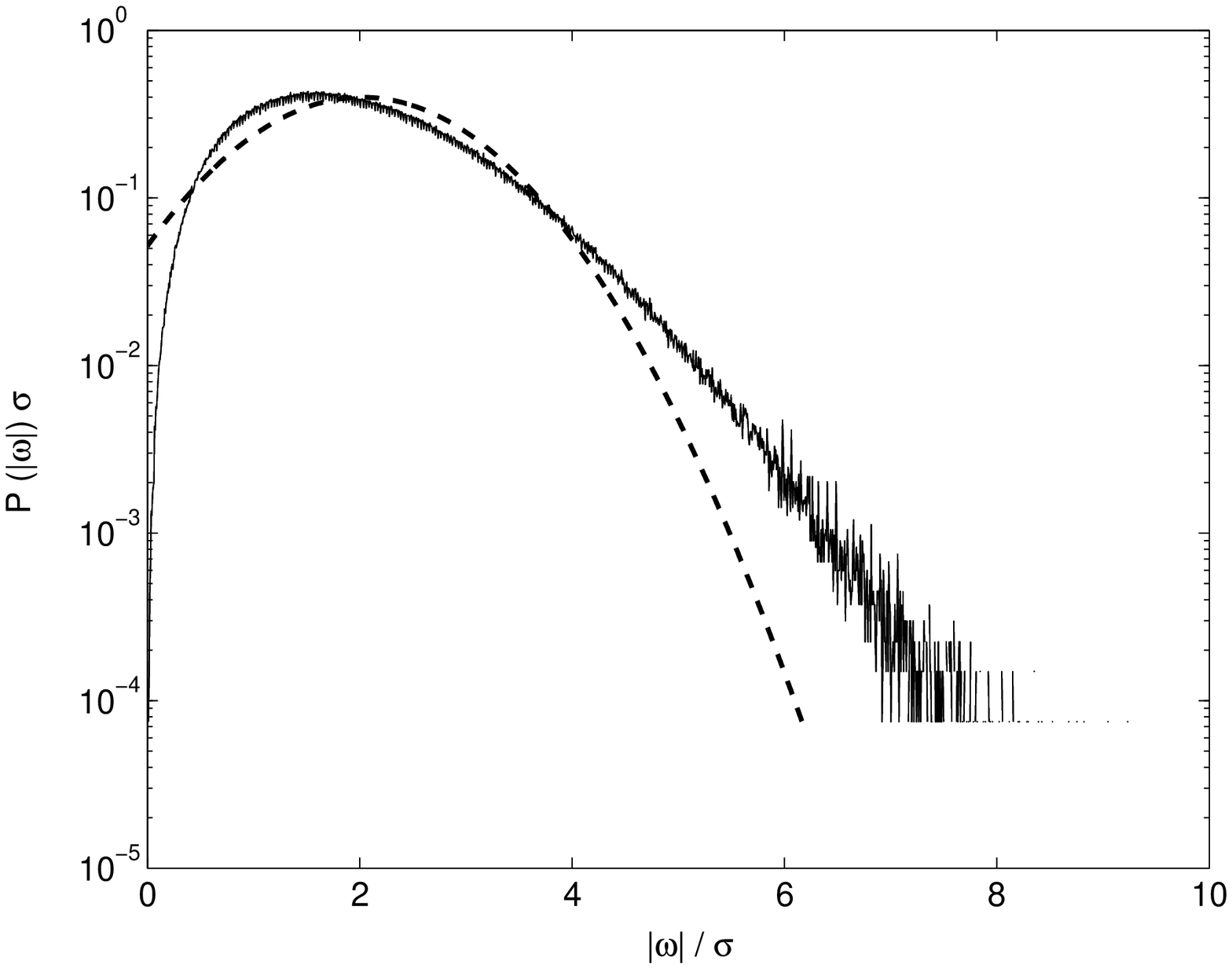}
\includegraphics[height=2.0in]{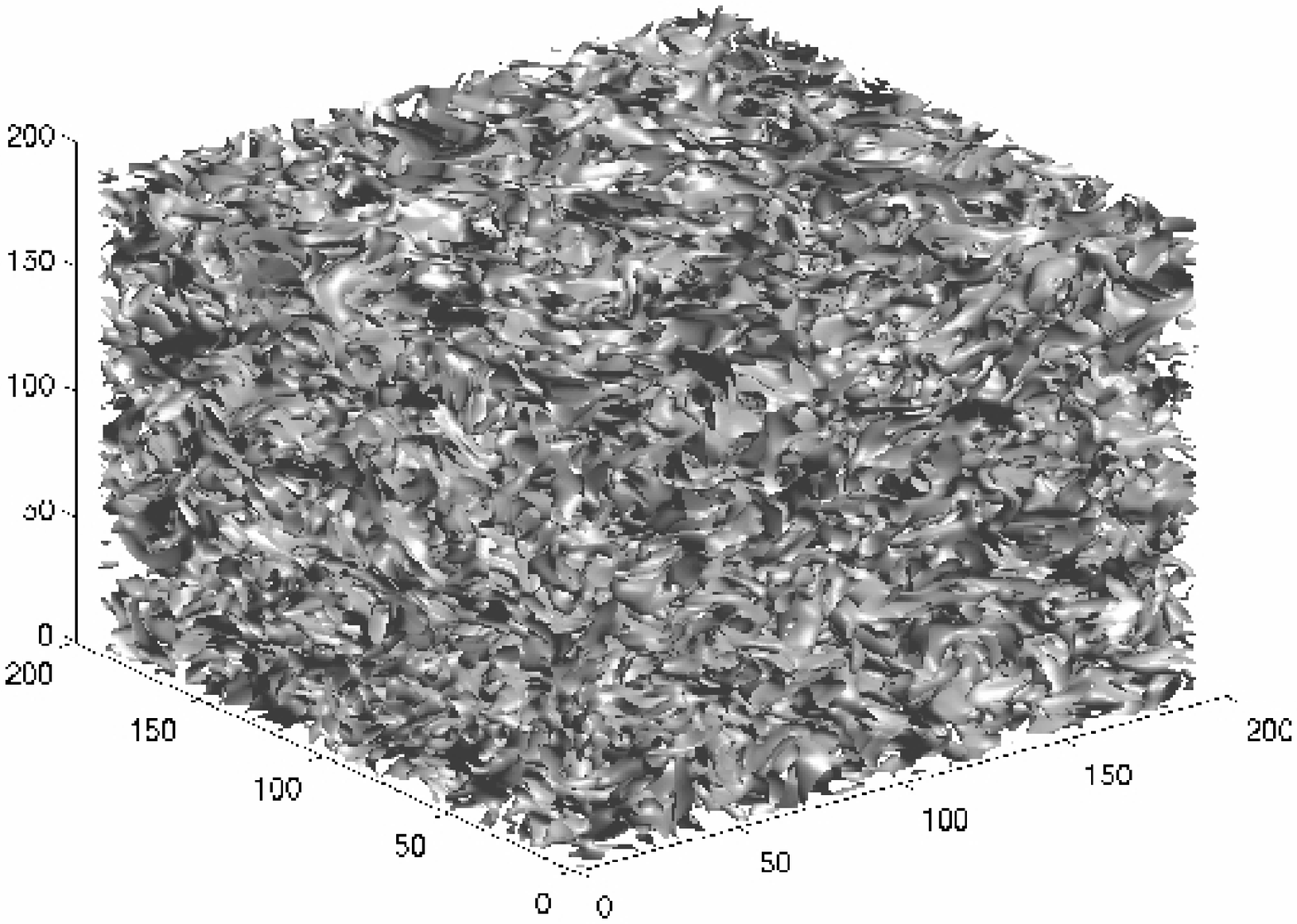}
\caption{\label{vortq1}(a) Semilog plot of the normalized probability 
distribution 
$P(|\omega|)$ of $|\omega|$ ($\omega$ is the vorticity) at 
dimensionless time $\tau=\tau_c$, with $E(k,t_0)\sim k$. The 
dashed-line plot is a normalized Gaussian distribution for comparison.\\
(b) Plot of iso-$|\omega|$ surfaces for 
the isovalue $|\omega|=\langle|\omega|\rangle$ at $\tau=\tau_c$, with 
$E(k,t_0)\sim k$.}
\end{figure}

\begin{figure}
\includegraphics[height=2.0in]{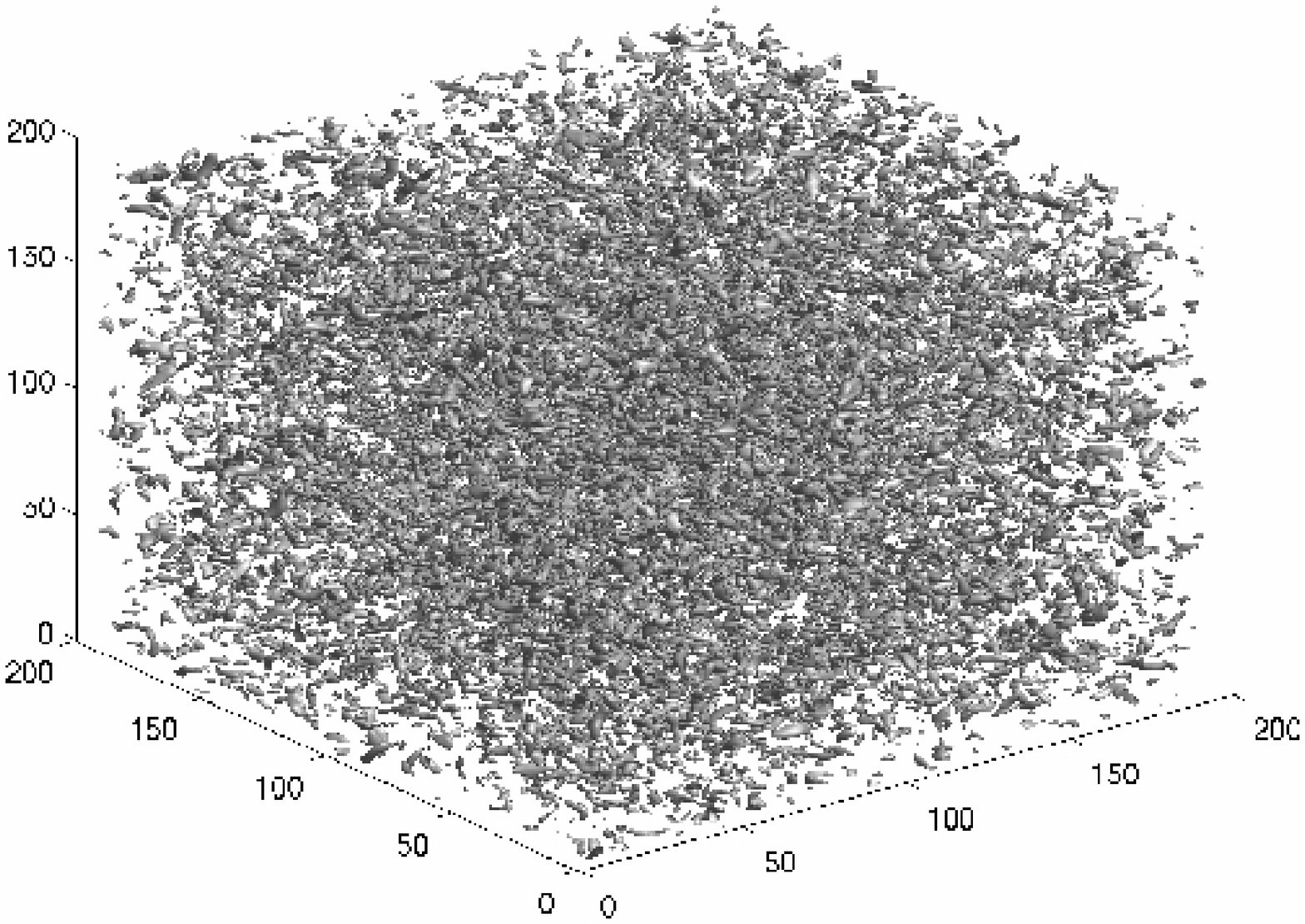}
\includegraphics[height=2.0in]{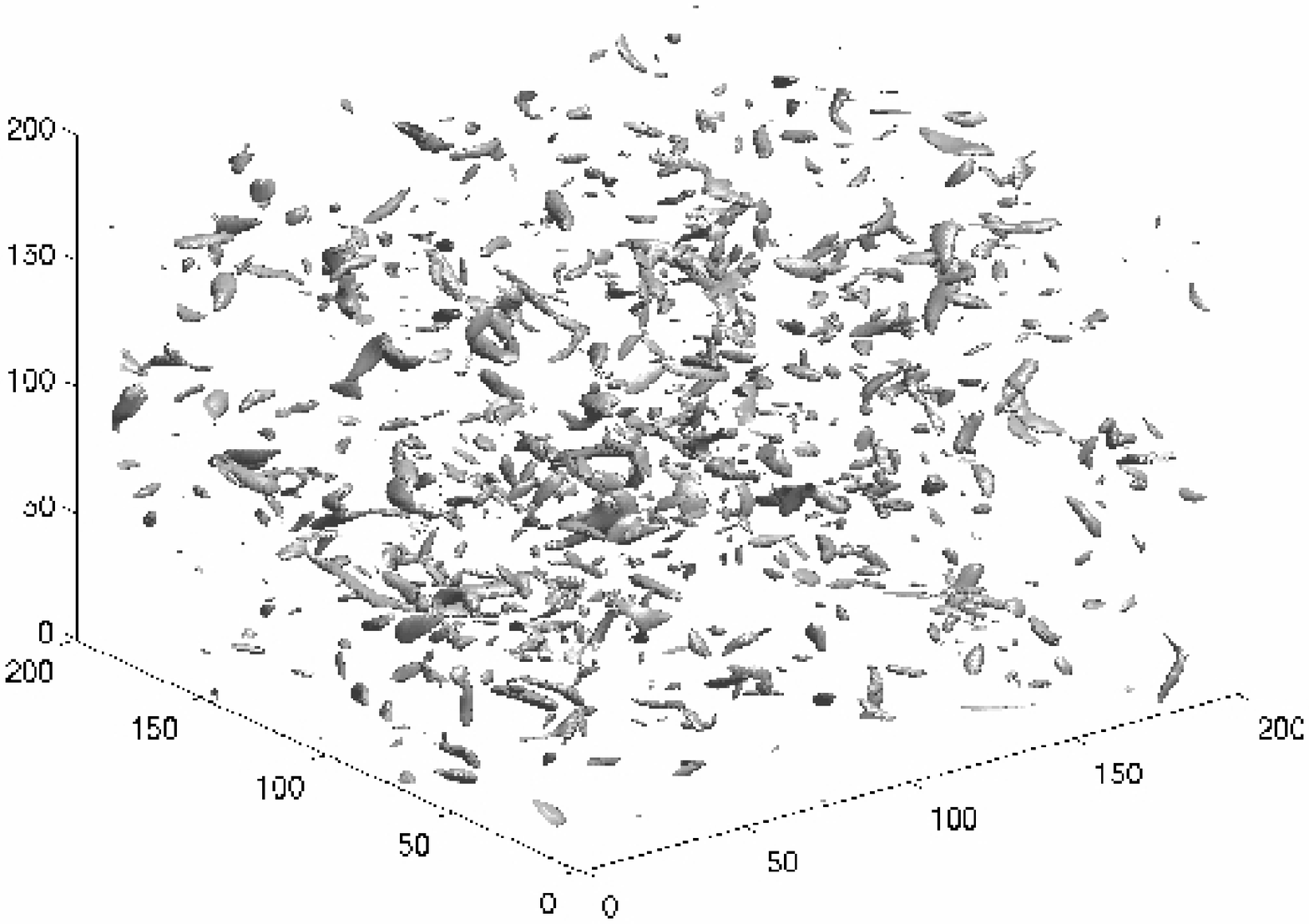}
\caption{\label{vortmaxq1}(a) Plot of iso-$|\omega|$ surfaces for 
the isovalue $|\omega|=\langle|\omega|\rangle+2\sigma$ ($\sigma$ is the 
standard deviation), at dimensionless time $\tau<<\tau_c$, with 
$E(k,t_0)\sim k$.\\(b) Plot of iso-$|\omega|$ surfaces for 
the isovalue $|\omega|=\langle|\omega|\rangle+2\sigma$ at $\tau=\tau_c$, with 
$E(k,t_0)\sim k$.} 
\end{figure}

In Fig. \ref{vortq1}(a), we plot the normalized probability 
distribution $P(|\omega|)$ of $|{\bf\omega}|$ ($\omega$ is the vorticity) at 
$\tau=\tau_c$. The distribution peaks at 
$|\omega|=1.5\sigma$ as in Fig. \ref{vort}(a), however it does not exhibit 
a stretched-exponential tail. In Fig. 
\ref{vortq1}(b), we plot iso-$|{\bf\omega}|$ 
surfaces for the isovalue $|\omega|=\langle|\omega|\rangle$ at $\tau=\tau_c$, 
which appear to be shredded sheet-like structures (observed throughout the 
isovalue range 
$[\langle|\omega|\rangle-\sigma,\langle|\omega|\rangle+\sigma]$). At early 
times $\tau<<\tau_c$, regions of intense vorticity (with the isovalue 
$|\omega|=\langle|\omega|\rangle+2\sigma$) were found to be 
structure-less `blobs' of length of the order of the grid spacing (see Fig. 
\ref{vortmaxq1}(a)), in contrast to the sheet-like structures in Fig. 
\ref{vortmax}(a). At $\tau=\tau_c$, isosurfaces of intense vorticity (in Fig. 
\ref{vortmaxq1}(b)), appear to be roughly ellipsoidal in shape with semiaxes 
of the order of the grid spacing and differ markedly from the 
coherent filaments in the `cascade-type' case shown in Fig. 
\ref{vortmax}(b). 
\subsection{Strain-Rate Tensor}
\begin{figure}
\includegraphics[height=2.0in]{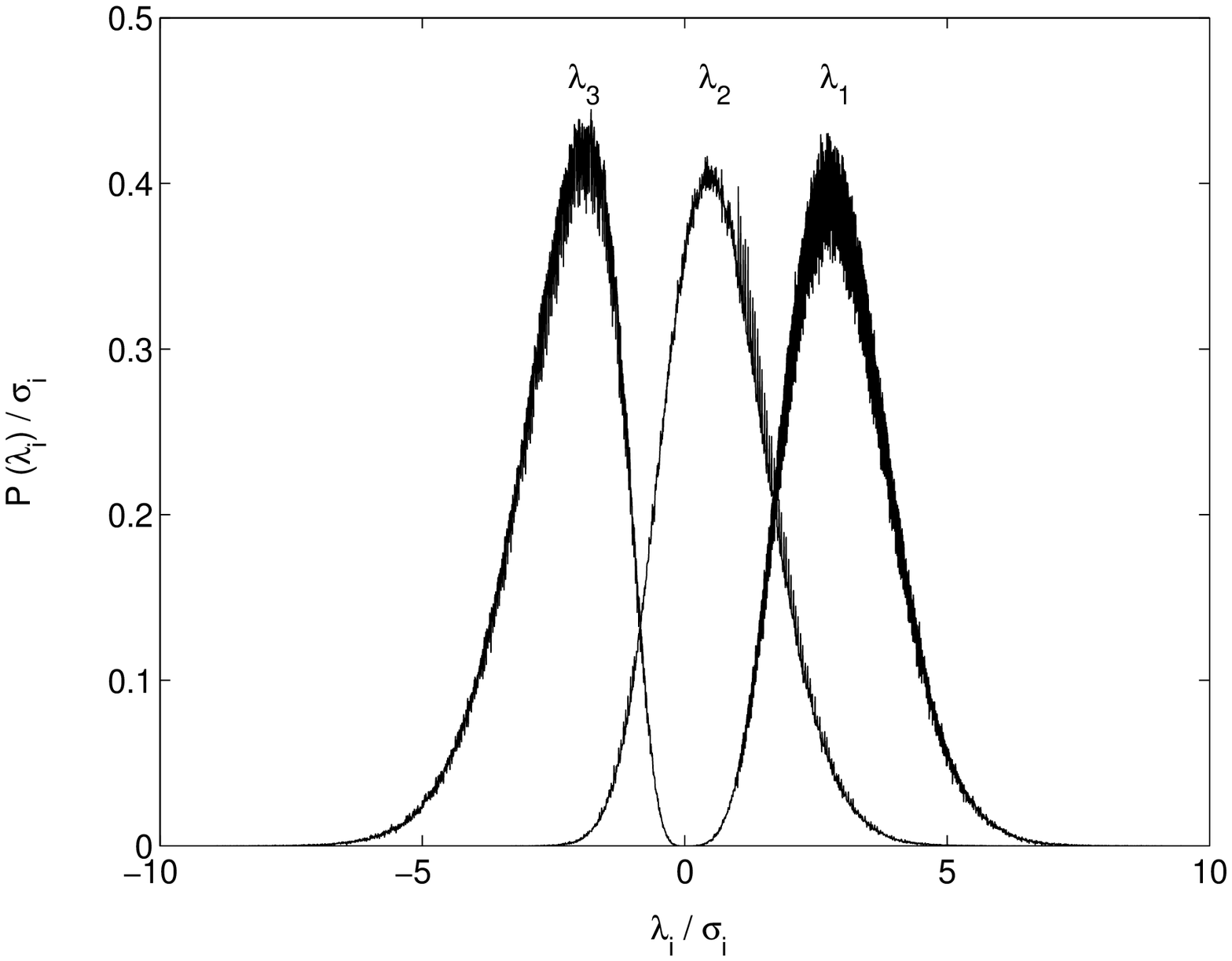}
\includegraphics[height=2.0in]{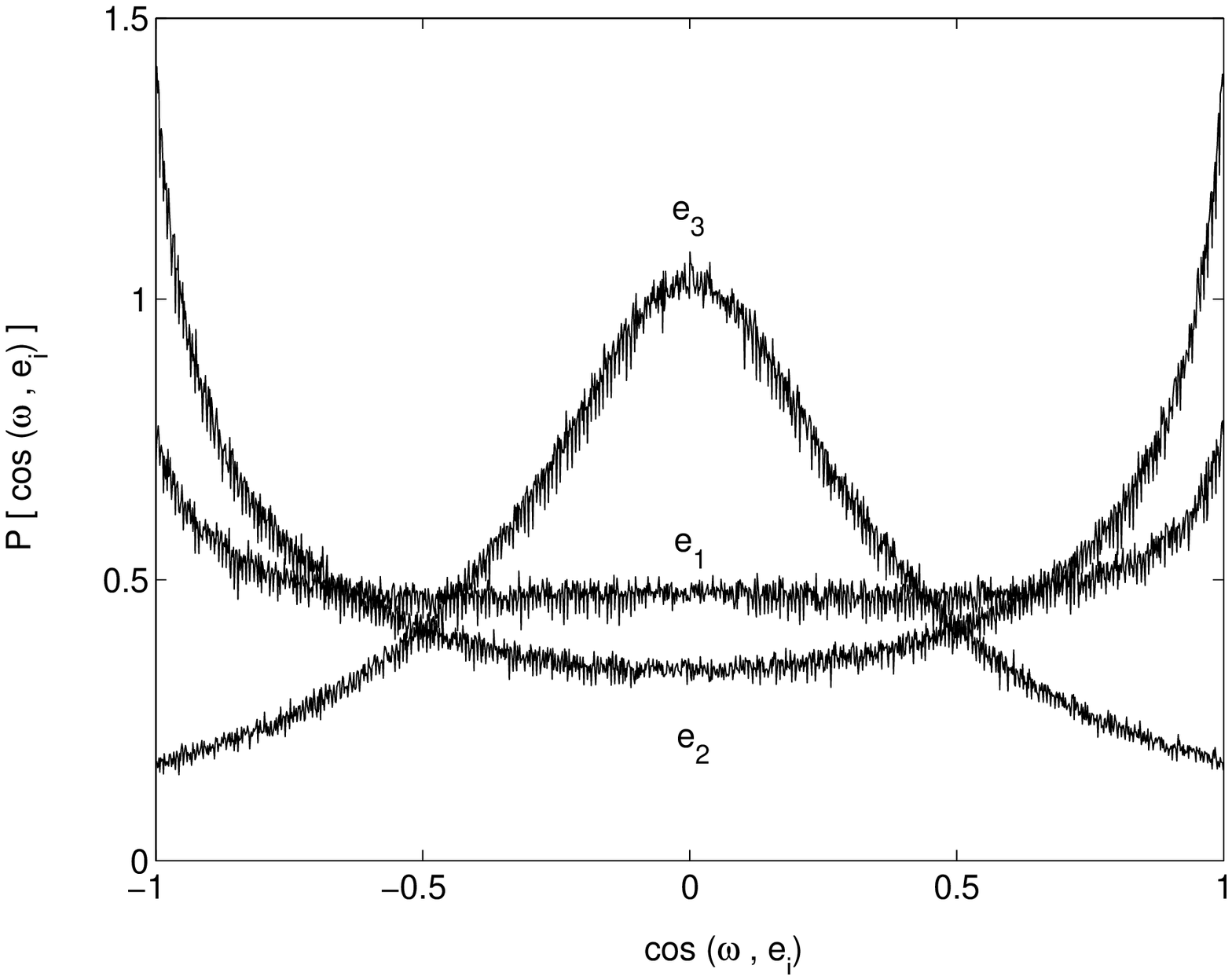}
\caption{\label{eigq1}(a) Plot of the normalized probability distribution
$P(\lambda_i)$ of the eigenvalues $\lambda_i$ of the strain-rate tensor 
$S_{ij}$, $i,j=1,2,3$, at 
dimensionless time $\tau=\tau_c$, with $E(k,t_0)\sim k$.\\
(b) Plot of the normalized probability distribution of cosine of 
the angle between $\omega$ and the eigenvectors $e_i$ of $S_{ij}$ at 
$\tau=\tau_c$, with $E(k,t_0)\sim k$.}
\end{figure}

In Fig. \ref{eigq1}(a), we plot the normalized probability 
distribution $P(\lambda_i)$ of the eigenvalues 
$\lambda_i$ of the strain-rate tensor $S_{ij}$ at $\tau=\tau_c$. We find 
that $\lambda_2$ has a positive mean. 
The statistically preferred ratio of the mean eigenvalues 
$\langle\lambda_1\rangle:\langle\lambda_2\rangle:\langle\lambda_3\rangle$ was 
found to equal $4.6:1:-5.6$, which differs only marginally from that 
obtained in 
the `cascade-type' case. In Fig. 
\ref{eigq1}(b), we plot the normalized probability distribution 
of cosine of the angle between $\omega$ and the 
eigenvectors $e_i$ of $S_{ij}$ at $\tau=\tau_c$. 
As in Fig. \ref{eig}(b), we observe that $\omega$ is preferentially parallel 
(or antiparallel) to $e_2$ (with a reduced peak-to-valley ratio $\approx4.1$) 
and perpendicular to $e_3$, while no specific 
angular relationship is observed relative to $e_1$ (though we note a small 
probability of alignment at cos($\omega,e_1$)=$\pm1$).
\begin{figure}
\includegraphics[height=2.0in]{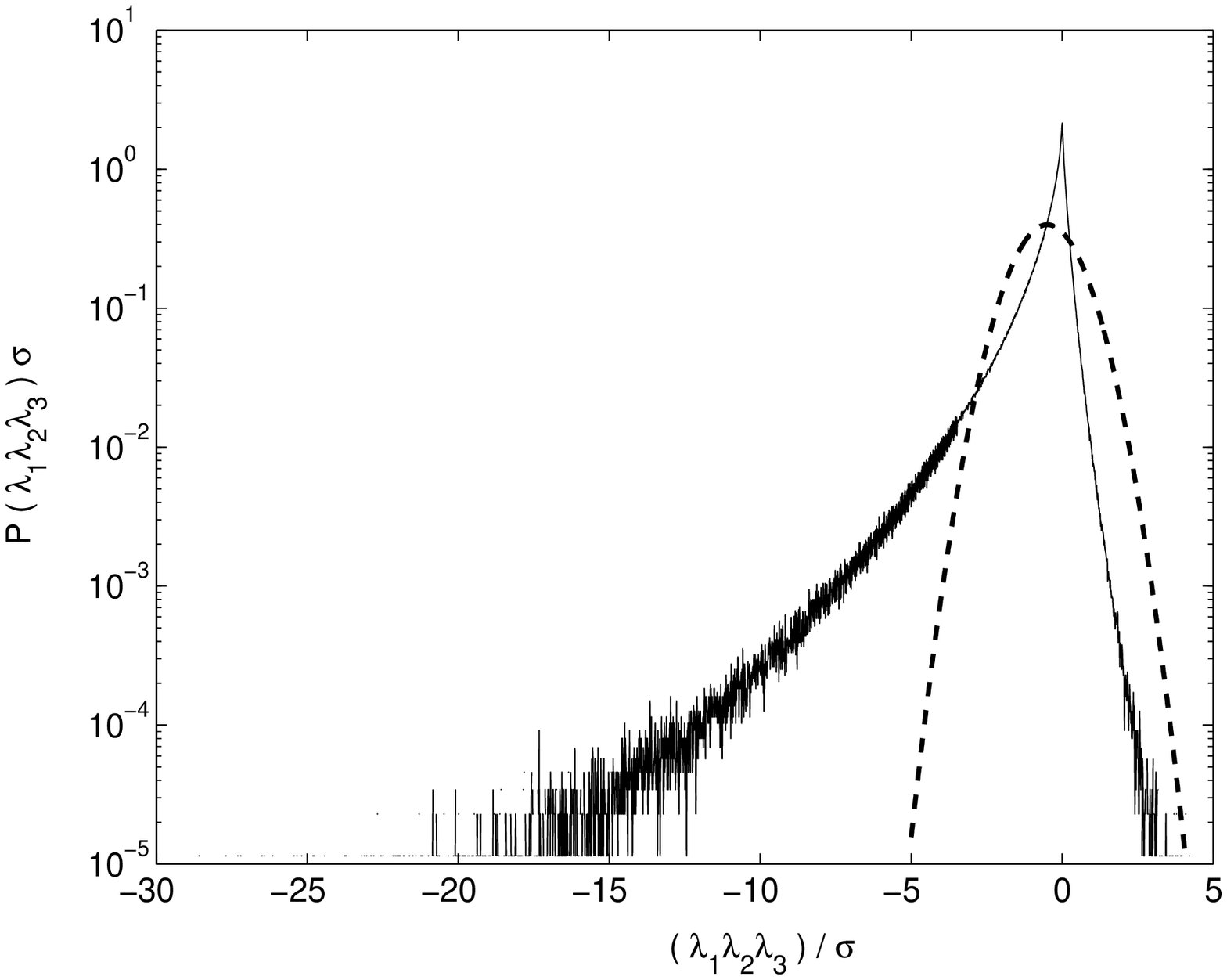}
\includegraphics[height=2.0in]{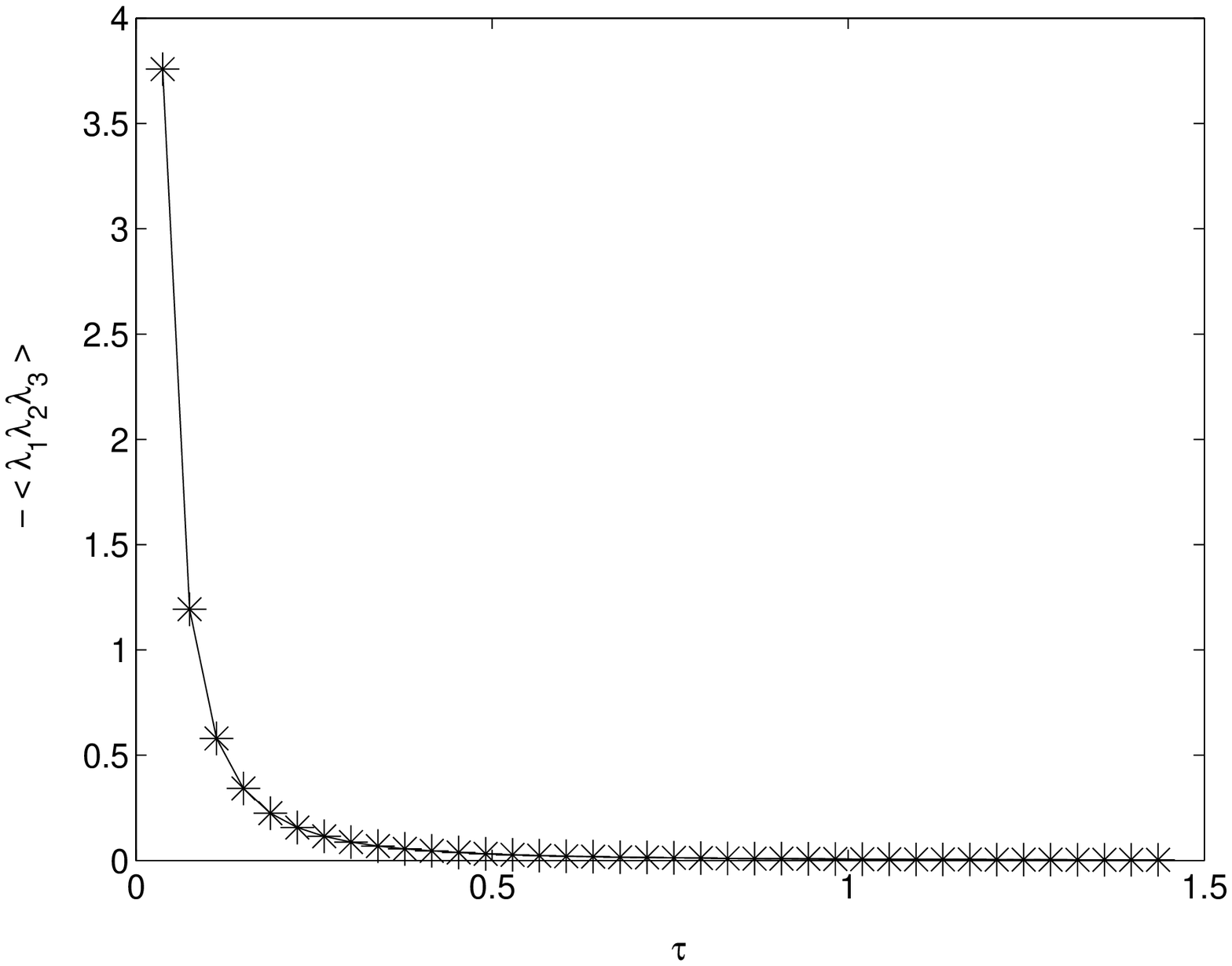}
\includegraphics[height=2.0in]{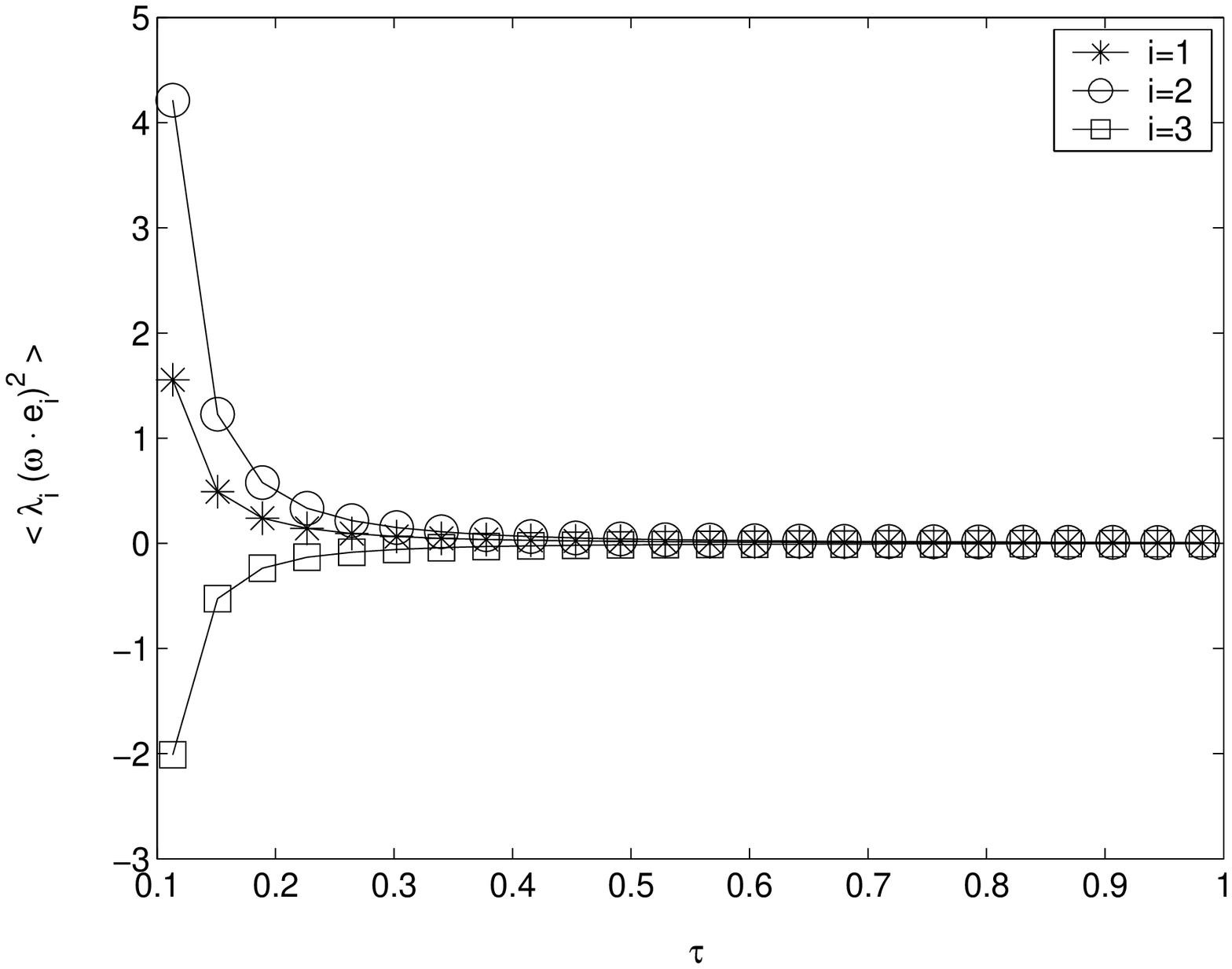}
\caption{\label{betchovq1}(a) Semilog plot of the normalized probability 
distribution 
$P(\lambda_1\lambda_2\lambda_3)$ of $\lambda_1\lambda_2\lambda_3$, with 
$E(k,t_0)\sim k$. The dashed-line plot is 
a normalized Gaussian distribution for comparison.\\
(b) Plot of $-\langle\lambda_1\lambda_2\lambda_3\rangle$ (see Eq. 
(\ref{betchoveqn})) as a 
function of the dimensionless time $\tau$, with $E(k,t_0)\sim k$.\\
(c) Plot of the rates of production of the 
mean enstrophy $\langle\lambda_i(\omega\cdot e_i)^2\rangle$ along the 
eigenvectors $e_i$ of $S_{ij}$, as a function of the 
dimensionless time $\tau$, with $E(k,t_0)\sim k$.}
\end{figure}
\begin{figure}
\includegraphics[height=2.0in]{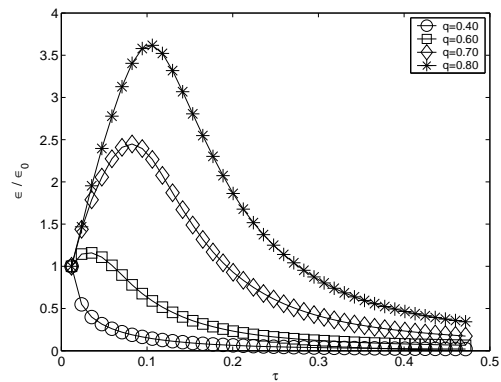}
\caption{\label{crossover}Plot of the normalized kinetic energy-dissipation 
rate $\epsilon(\tau)/\epsilon_0$ as a function of the dimensionless time 
$\tau$, with $E(k,t_0)\sim k^qe^{-k^q}$, $0.4\le q\le0.8$.} 
\end{figure}
\subsection{Mean Enstrophy Production Rate}
In Fig. \ref{betchovq1}(a), we plot the normalized probability distribution 
$P(\lambda_1\lambda_2\lambda_3)$ of $\lambda_1\lambda_2\lambda_3$ at 
$\tau=\tau_c$. 
As in Fig. \ref{betchov}(a), we find that the distribution has a negative 
mean (a relatively smaller skewness equal to $-3.90$ was found). 
In Fig. \ref{betchovq1}(b), 
we plot the value of $-\langle\lambda_1\lambda_2\lambda_3\rangle$ (see 
Eq. (\ref{betchoveqn})) as a function of the dimensionless time $\tau$ and 
find that it decreases monotonically, in contrast to 
Fig. \ref{betchov}(b). In Fig. 
\ref{betchovq1}(c), we plot $\langle\lambda_i(\omega\cdot e_i)^2\rangle$, 
the rates of production of the mean enstrophy along eigenvectors $e_i$ 
as a function of the dimensionless time $\tau$, and find that the production 
rates {\it decrease} monotonically along all the principal axis 
directions, in remarkable contrast to 
Fig. \ref{betchov}(c). 
In Fig. \ref{betineq}, we plot inequality (\ref{betchovineq}) 
as a function of the 
dimensionless time $\tau$, and find that the inequality monotonically 
approaches 
the same asymptotic value $\approx0.57$ as in the `cascade-type' case.
\subsection{Crossover}
From Figs. \ref{prelim}(c) and \ref{prelimq1}(c), it is evident that some 
representative set of initial energy spectra may be utilised to exhibit a 
crossover from non-`cascade-type' to `cascade-type' behaviour, with the 
kinetic energy-dissipation rate serving as a diagnostic. In particular, 
one may choose the set $E(k,t_0)\sim k^qe^{-k^q}$ with parameter $q$. 
In Fig. \ref{crossover}, we plot $\epsilon(\tau)/\epsilon_0$ as a function 
of the dimensionless time $\tau$ (with $\tau_0$ calculated using the initial 
energy spectrum with $q=0.4$). On 
varying $q$ in the range 
$[0.4,0.8]$, we find that $\epsilon(\tau)/\epsilon_0$ exhibits a smooth 
crossover at $q\approx0.6$ from non-`cascade-type' to `cascade-type' behaviour. 
\section{Conclusions}
To summarize, we have presented results from a systematic numerical study of 
structural properties of an unforced, incompressible, homogeneous, and 
isotropic turbulent fluid with 
an initial energy spectrum that develops a cascade to large wavenumbers. 
The results are contrasted with those from the power-law initial spectra 
investigated in Refs. \cite{Kalelkar,Ditlevsen}, which do not exhibit such a 
cascade. Differences are noted in plots of vorticity isosurfaces, the 
temporal evolution of the kinetic energy-dissipation 
rate, and the rates of production of the mean enstrophy along the principal 
axes of the strain-rate tensor. However, preferential alignment of the 
vorticity vector with 
the intermediate eigenvector of the strain-rate tensor is found to be 
unchanged. A crossover between 
non-`cascade-type' and `cascade-type' behaviour is shown numerically for a 
specific set of initial energy spectra. Our study shows that decaying 
turbulence is strikingly different for different types of initial conditions. 
Initial conditions that lead to a cascade yield features similar to those 
obtained in the statistically steady case. However, power-law initial 
conditions that do not lead to an energy cascade exhibit several qualitatively 
new features which we have described above.
\begin{acknowledgments}
The author thanks T. Kalelkar, R. Pandit, and R. Govindarajan for discussions, 
D. Mitra for the code, SERC (IISc) for computational resources, and CSIR 
(India) for financial support. 
\end{acknowledgments}


\begin{thebibliography}{11}
\bibitem{smallfn} In Ref. \cite{Jimenez} it has been shown that filamentary 
structures in regions of intense vorticity contain only a small percentage 
of the total enstrophy.
\bibitem{hosfn} In Ref. \cite{Hosokawa}, filamentary structures are visualized 
in a $512^3$ DNS of decaying turbulence.
\bibitem{modefn} We note that in Refs. \cite{Ashurst}, the strain-rate 
eigenvalue ratio was found to equal $3:1:-4$ based on the most probable 
values, as opposed to the mean.
\bibitem{Redfn} The analogous equation for the mean-squared strain-rate 
$\langle s^2\rangle\equiv\langle S_{ij}S_{ij}\rangle$ is redundant since 
$\langle s^2\rangle=\langle\omega^2\rangle/2$ in homogeneous, incompressible 
flows. The probability distribution $P(|s|)$, $|s|\equiv\sqrt{S_{ij}S_{ij}}$ 
was found to resemble $P(|\omega|)$.
\bibitem{Refn} At dimensionless time $\tau=\tau_c$, the Reynolds 
number $Re$ were found to be of the same order of magnitude in 
both `cascade-type' and `power-law' cases.
\end{thebibliography}
\end{document}